\documentclass{emulateapj}
\usepackage{apjfonts}
\usepackage{bm}
\usepackage{amsmath}

\newcommand\lsim{\mathrel{\rlap{\lower4pt\hbox{\hskip1pt$\sim$}}
        \raise1pt\hbox{$<$}}}
\newcommand\gsim{\mathrel{\rlap{\lower4pt\hbox{\hskip1pt$\sim$}}
        \raise1pt\hbox{$>$}}}
\newcommand{\D}{\mathrm{d}}

\newcommand{\m}{\mathrm{m}}

\newcommand{\Hz}{\mathrm{Hz}}
\newcommand{\GHz}{\mathrm{GHz}}

\newcommand{\dm}{\mathrm{dm}}
\newcommand{\gas}{\mathrm{gas}}
\newcommand{\vir}{\mathrm{vir}}
\newcommand{\mx}{\mathrm{max}}
\newcommand{\mK}{\mathrm{mK}}
\newcommand{\muK}{\mu\mathrm{K}}
\newcommand{\OmegaB}{\Omega_\mathrm{B}}

\shorttitle{Virial Shock Detection in Galaxy Clusters with the Sunyaev-Zel'dovich Effect}
\shortauthors{Kocsis, Haiman, \& Frei}

\begin{document}

\title{Can Virialization Shocks be Detected Around Galaxy Clusters \\ Through the Sunyaev-Zel'dovich Effect?}

\author{Bence Kocsis}
\affil{Institute of Physics, E\"otv\"os University, P\'azm\'any P. s. 1/A,
1117 Budapest, Hungary; bkocsis@complex.elte.hu}
\author{Zolt\'an Haiman}
\affil{Department of Astronomy, Columbia University, 550 West 120th Street,
New York, NY 10027; zoltan@astro.columbia.edu}
\author{Zsolt Frei}
\affil{Institute of Physics, E\"otv\"os University, P\'azm\'any P. s. 1/A,
1117 Budapest, Hungary; frei@alcyone.elte.hu}

\hspace{\baselineskip}

\begin{abstract}

In cosmological structure formation models, massive non--linear
objects in the process of formation, such as galaxy clusters, are
surrounded by large-scale shocks at or around the expected virial
radius. Direct observational evidence for such virial shocks is
currently lacking, but we show here that their presence can be
inferred from future, high resolution, high--sensitivity observations
of the Sunyaev-Zel'dovich (SZ) effect in galaxy clusters.  We study
the detectability of virial shocks in mock SZ maps, using simple
models of cluster structure (gas density and temperature
distributions) and noise (background and foreground galaxy clusters
projected along the line of sight, as well as the cosmic microwave
background anisotropies).  We find that at an angular resolution of
$2''$ and sensitivity of $10\muK$, expected to be reached at $\sim
100$ GHz frequencies in a $\sim 20$hr integration with the forthcoming
ALMA instrument, virial shocks associated with massive ($M\sim
10^{15}~{\rm M_\odot}$) clusters will stand out from the noise, and
can be detected at high significance.  More generally, our results
imply that the projected SZ surface brightness profile in future,
high--resolution experiments will provide sensitive constraints on the
density profile of cluster gas.

\end{abstract}



\section{Introduction}

In cosmological theories of structure formation, non--linear objects
form when overdense dark matter perturbations turn around, collapse,
and settle into virial equilibrium (e.g. \citealt{peebles93} and
references therein). Gas initially collapses together with the dark
matter, but eventually encounters nearly stationary material that had
already collapsed. Since the gas is falling in at supersonic
velocities, it is slowed down by hydrodynamical shocks, and these
shocks are thought to heat the gas to the virial temperature of the
dark matter halo.

In spherically symmetric models, and in the absence of dissipation, a
single strong gaseous shock occurs at approximately half of the
turn--around radius \citep{bert85}, coinciding with the ``virial
radius'' of the dark matter halo.  More realistically, the behavior of
the post--shock gas depends sensitively on its cooling time
\citep{ro77}.  On galactic scales ($M\sim 10^{12}~{\rm M_\odot}$) and
below, and increasingly toward high redshifts ($z\gsim 3$), the gas
can cool rapidly and loose its pressure support, and hence continue
its infall.  On these scales, the existence of large--scale shocks
have been recently called into question by models in which the bulk of
the infalling gas remains cold, and reaches the central regions of the
dark halo before encountering shocks \citep{bd03, ketal04}.  On larger
scales, however, where cooling times are long, such as for galaxy
clusters, the existence of virial shocks remains an unambiguous
prediction of cosmological structure formation theories. Detailed
three--dimensional hydrodynamical simulations of cluster formation
(e.g. \citealt{evrard90, bn98}) have confirmed the existence of virial
shocks, with strong discontinuities in gas density and
temperature. These and subsequent simulations have also revealed that
the infall is anisotropic, with gas falling in to the cluster
potential along cosmic filaments.  As a result, the radial location
and strength of the shocks varies along different directions.

The virial shocks are a fundamental ingredient of cosmic structure
formation, and may be responsible for diverse phenomenae, such as
generating large--scale cosmic magnetic fields \citep{Bagchi}
and accelerating electrons to contribute to the diffuse cosmic
gamma--ray background \citep{lw00}.  The radial location of the
shocks, in principle, also contains information on the cosmological
parameters \citep{vhs02}.  Despite their importance, direct evidence
for the existence of such shocks does not yet exist.  The major
difficulty in observing the virial shock is that it is expected to lie
several Mpc (and several arcminutes) away from the cluster center, a
location at which signals such as the X--ray surface brightness
\citep{tozzi00}, or galaxy number density/peculiar velocities (which
could reveal density caustics, \citealt{rines}) diminish rapidly.

In this paper, we consider the detectability of virial shocks in
future observations of galaxy clusters through the Sunyaev-Zel'dovich
(SZ) effect.  The thermal SZ effect is a secondary distortion of the
cosmic microwave background (CMB) spectrum caused by the hot
intra--cluster gas along the line of sight to the surface of last
scattering (see \citealt{sz80} for a review).  The cool CMB photons
undergo inverse Compton scattering on the hot electrons, gaining on
average a small amount of energy in the process, creating an intensity
decrement at low frequencies ($\nu\lesssim 218\GHz$) and an increment
at high frequencies. The SZ effect is the dominant source of CMB
anisotropy at small angular scales.

The SZ effect has recently become a valuable observational tool
\citep{Birkinshaw}. Several programs have begun to map out massive
clusters of galaxies, study the intracluster medium (ICM), and
constrain cosmological parameters.  Current instruments are now
detecting and imaging clusters at high signal-to-noise, and the next
generation of instruments should be capable of mapping significant
portions of the sky as a means of finding clusters of galaxies (see
\citealt{carl02} for a review). Several studies have predicted the
number of clusters that could be detected in future SZ surveys
\citep{Bartlett, Holder1, hmh01, Barbosa, Kneissl}. The survey yields
are quite impressive. Next generation instruments, such as the Atacama
Cosmology Telescope (ACT), South Pole Telescope (SPT), and the {\it
Planck} satellite\footnote{See www.hep.upenn.edu/act,
astro.uchicago.edu/spt, and www.rssd.esa.int/index.php?project=PLANCK,
respectively.}, are expected to detect several clusters per day; the
large resulting samples can be used to select the most massive and
most regular clusters that will be best suited for the studies
proposed here.

The SZ effect is ideally suited to study the ``outskirts'' of
clusters, because the SZ temperature decrement profile is relatively
flat (e.g. $\Delta T \propto \rho$, whereas the X--ray emission is
proportional to the square of the local density; \cite{KS01}).
Although our main focus is to assess the significance at which the
shocks can be detected, we also consider the more general problem of
constraining the cluster gas distribution, as well as the structure of
the dark matter halos themselves.

The detection of sharp features, such as the virial shocks, calls for
high sensitivity, high--resolution maps of the SZ surface brightness
profile of the most massive clusters.  For this reason, we here focus
on predictions appropriate for the Atacama Large Millimeter Array
(ALMA)\footnote{See www.alma.nrao.edu.}, a telescope array expected
to be operational in 2012 and deliver $\sim$arcsecond resolution,
high--sensitivity imaging of clusters. Our results can be scaled to
apply to other instruments with different parameters.

This paper is organized as follows.
In \S~\ref{sec:alma}, we summarize the relevant characteristics of ALMA.
In \S~\ref{sec:rhoprofiles}, we describe our models for the structure
of galaxy clusters.  These models are based on standard descriptions
of gas in hydrostatic equilibrium with a dark matter halo, except that
we introduce additional free parameters that allow us to vary the
location and sharpness of the 'virial shock.
In \S~\ref{sec:szprofiles}, we compute and contrast the SZ surface
brightness profiles in models with different virial shocks.
In \S~\ref{sec:noise}, we discuss the sources of noise in the SZ
surface brightness maps.
In \S~\ref{sec:edgedetection}, we present simple estimates to argue
that, in the face of noise, smooth cluster profiles can be
distinguished at high significance from profiles that include a virial
shock.
In \S~\ref{sec:parameters}, we go one step further, and compute the
statistical accuracy at which the location and sharpness of the shock
fronts, as well as other parameters describing the cluster profile,
can be determined from future SZ maps.
Finally, in \S~\ref{sec:conclusions} we summarize our conclusions and
discuss the implications of this work.  Several technical points,
related to our statistical analysis, are discussed in the Appendices.

Throughout this paper we assume a standard cold--dark matter cosmology
($\Lambda$CDM), with ($\Omega_\Lambda$, $\Omega_M$, $\Omega_b$, $H_0$)
= (0.7, 0.3, 0.045, 70 km s$^{-1}$ Mpc$^{-1}$), consistent with the
recent results from {\it WMAP} \citep{spergel03}.


\section{ALMA}
\label{sec:alma}

The three crucial characteristics of any instrument for the detection of
the virial shocks are angular resolution (needed since the features
caused by the shocks are sharp), angular coverage (needed since the
features can extend coherently over arcminute scales), and sensitivity
(the surface brightness change across the shock--front is as small as
a few $\mu$K).  ALMA has an ideal combination of these three
qualities, and is well suited to studying the gas density cutoff.
ALMA will be comprised of sixty four 12-meter sub-millimeter quality
antennae, with baselines extending from 150 m up to 10 km.  Its
receivers will cover the range from $\nu=70$ to $900$ GHz.
Anticipated SZ temperature sensitivity for a $\Delta t_s \approx 60 s$
integration period is between $\sigma_D\approx 100\muK$ and $24$mK
for the compact configuration depending on the frequency of the
detection.

The expected aperture efficiency of an antenna has been taken from
\citet{ALMA1}.  Generally, a baseline $B$ corresponds to a beam radius
\begin{equation}
\Delta\phi=\frac{c}{\nu B}=
2"\times\left(\frac{\nu}{100\GHz}\right)^{-1}
\left(\frac{B}{300\m}\right)^{-1}.
\end{equation}

The resulting temperature sensitivities for the compact
configuration ($B=150\m$) is given in Table
\ref{tab:ALMA}.\footnote{Adopted from
http://www.alma.nrao.edu/info/sensitivities.}  In our calculations
below, we will assume a frequency of 100 GHz, but
Table~\ref{tab:ALMA} shows the required integration time and
baseline for several different frequencies.  The second and third
columns show the beam diameter and the sensitivity at a fixed
integration time of 60s. In the fourth and fifth columns, we
calculated the baseline needed to have $\Delta\phi\leq 2"$ and the
corresponding sensitivity at the same fixed integration time of
$60s$. In the last column we calculated the integration time
$\Delta t_{\rm int}$ needed to reduce the r.m.s. noise variance to
$10\muK$ (using the scaling $\sigma_D\propto\sqrt{B^2\Delta t_{\rm
int}}$).

\begin{deluxetable}{cccccc}
\tablecolumns{6} \tablewidth{0pt}
\tablecaption{\label{tab:ALMA}Temperature sensitivities and
integration times} \tablehead{ \colhead{$\nu$} & \colhead{$\phi$}
& \colhead{$(\sigma_D)_{60s}$} & \colhead{$B$} &
\colhead{$(\sigma_D)_{60s}$} & \colhead{$(\Delta t_{\rm
int})_{10\muK}$}
\\
\colhead{[GHz]} & \colhead{[arcsec]} & \colhead{[mK]} &
\colhead{[m]} & \colhead{[mK]} & \colhead{[hr]}} \startdata
      35      & 11.79   & 0.1079 &  884  & 3.747 & 67\\
      90      & 4.584   & 0.1871 &  344  & 0.983 & 31\\
      140     & 2.947   & 0.2518 &  221  & 0.547 & 23\\
      230     & 1.794    & 0.4317 & 150  & 0.432 & 31\\
      345     & 1.196   & 1.007  &  150  & 1.007 & 169\\
      409     & 1.009   & 1.799  &  150  & 1.799 & 539\\
      650     & 0.6347  & 13.67  &  150  & 13.67 & $3.11\times10^4$ \\
      850     & 0.4853  & 24.46  & 150  & 24.46 & $9.97\times 10^4$ \\
\enddata
\end{deluxetable}

For the present study we will assume that the sensitivity of the
detector is $\sigma_D\lesssim 10\mu$K, and its angular resolution is
$\Delta\phi\approx 2"$.  As Table~\ref{tab:ALMA} shows, these
assumptions are realistic for ALMA, requiring $23-170$ hr integration
for all frequencies below $350 \GHz$.  For even longer observation
times, the detector noise can be neglected compared to the statistical
fluctuations of astrophysical origin (the latter, which has a
magnitude of $\sim10\mu$K, will be discussed in \S~\ref{sec:noise}
below).


\section{Self-Similar Cluster Density and Temperature Profiles}
\label{sec:rhoprofiles}

In this section, we describe the models we adopt for the density and
temperature profiles of gas in the cluster. These models assume that
clusters are described by spherically symmetric dark matter halos.  We
first follow \cite{KS01} to obtain self--similar gas density and
temperature profiles. These profiles are then truncated and normalized
according to the assumed location and sharpness of a virial
shock.

Many high-resolution N-body simulations suggest that the dark matter
density profile $\rho_\dm(r)$ is well described by a self--similar
form: $\rho_\dm(r)=\rho_s y_\dm(r/r_s)$ where $\rho_s$ is the mass
density normalization factor, $r_s$ is a length scale, and $y_\dm(x)$
is a non-dimensional function representing the profile. The density
normalization, $\rho_s$, is determined to yield mass $M_\vir$ when
$\rho_\dm(r)$ is integrated within the virial radius, $r_\vir$. The
length scale $r_s$ is defined as $r_s=r_\vir/c$, where $c$ is the
concentration parameter. We use a fitting formula from detailed gas
dynamical simulations for the concentration parameter (see
eq. \ref{eq:c} below), and calculate $r_\vir(M_\vir,z)$ from the
top--hat collapse model (see discussion below).

The dark-matter profile is approximated by the following analytic form
\begin{equation}\label{eq:y_dm}
y_\dm(x)=\frac{1}{x^\alpha(1+x)^{3-\alpha}},
\end{equation}
where the parameter $\alpha$ is assumed to be either $\alpha=1$ or
$\alpha=3/2$ (see \citealt{NFW97}, \citealt{moore99}, and
\citealt{jing00}).

The gas density profile can be obtained using a polytropic model in
hydrostatic equilibrium with the dark matter background. In this case,
the gas density and temperature profiles assume a self--similar form,
\begin{align}
\rho_\gas(r/r_s)&=\rho_\gas(0)y_\gas(r/r_s)\\
\label{eq:temperature}
T_{\gas}(r/r_s)&=T_{\gas}(0)y_{\gas}^{\gamma-1}(r/r_s),
\end{align}
where $P_{\gas}\propto\rho_{\gas}T_{\gas}\propto\rho^\gamma$ has
been used. The hydrostatic equilibrium equation
\begin{equation}
\rho_\gas^{-1}\frac{\D P_\gas}{\D r}= - G \frac{M(\leq r)}{r^2}
\end{equation}
can be solved \citep{suto98} for $y_\gas(x)$ for a fixed dark matter
mass distribution.
\begin{align}\label{eq:y_gas}
y_\gas^{\gamma-1}(x)=
1-3\eta^{-1}\frac{\gamma-1}{\gamma}\frac{c}{m(c)}\int_0^x\D
u\frac{m(u)}{u^2},
\end{align}
where $\eta$ is an integration constant, $x$ is defined as
\begin{equation}
x= \frac{r}{r_s},
\end{equation}
and $m(u)$ is the dimensionless mass within a distance $u$ from the
center
\begin{equation}\label{eq:m(x)}
m(x)\approx m_\dm(x)=4\pi\int_0^x\D u u^2 y_\dm(u).
\end{equation}
These integrals can be evaluated analytically \citep{suto98} for the
particular cases $\alpha=1$ and $\alpha=3/2$.

The mass-temperature normalization is given by the virial theorem
\begin{equation}\label{eq:mass-temperature}
\eta^{-1}=\frac{G\mu m_\mathrm{p}M_\vir}{3r_\vir k_\mathrm{B}
T_\gas(0)}.
\end{equation}

Both theoretical and numerical studies assert that the gas density
profile traces the dark matter density profile in the outer regions of
the halo. Therefore, the slopes of these two profiles are assumed to
track each other closely for $c/2<x<2c$.  This requirement fixes the
polytropic index $\gamma$, and the normalization $\eta$ (see equations
22, 23, and 25 in \citealt{KS01}).

The concentration parameters, $c$ and $c_{NFW}$ can be written in
terms of the virial mass and redshift for a given cosmological model.
A fitting formula based on numerical simulations \citep{eke01} is
\begin{align}\label{eq:c}
c(M_\vir,\alpha,z)&=\left\{
\begin{array}{c}
  c_{NFW} \text{~if~~} \alpha=1 \\
  c_{NFW}/1.7 \text{~if~~} \alpha=1.5 \\
\end{array}
\right.\\
c_{NFW}&=\frac{6}{1+z}\left(\frac{M_\vir}{10^{14}h^{-1}M_\odot}\right)^{-1/5}.
\end{align}
Equation~(\ref{eq:c}) supplies a one--to--one correspondence between
$c$ and $M_\vir$.

The self--similar model defined above does not assign a value for the
gas density normalization $\rho_\gas(0)$. Its value can be calculated
by requiring that the ratio of the total dark matter and gas mass
within some radius $x_\mx$ attain the universal average value of
$\Omega_\dm/\OmegaB$.  Simulations without feedback from galaxy
formation typically find values for the cluster gas mass fraction that
are only slightly lower than the input global baryon fraction
\citep{evrard97}.  Using cosmological parameters cited above
\begin{equation}\label{eq:rho_gas0}
\rho_\gas(0)=\rho_s\frac{m_\dm(x_\mx)}{m_\gas(x_\mx)}\frac{\OmegaB}{\Omega_\dm},
\end{equation}
where $m_\gas(x)$ is the accumulated dimensionless mass of the gas
analogous to $m_\dm(x)$ defined in equation~(\ref{eq:m(x)}). The
parameter $\rho_\gas(0)$ has only a mild $x_\mx$ dependence for
reasonably low $x_\mx$ values. Using a fixed $x_\mx=5c$ yields a
maximum error of 10\% for $x_\mx \in[c, 10c]$.

We calculate the virial radius, $r_\vir$, using the spherical top hat
collapse model, i.e. we take it to be the radius of a spherical region
with mean interior overdensity $\Delta_c$ relative the critical
(obtained from the fitting formula in equation 6 of \citealt{bn98})
that encloses a total mass $M_{\rm vir}$.  The corresponding angular
radius of the cluster is
\begin{equation}\label{eq:Theta}
\Theta_\vir=\frac{r_\vir}{d_A(z)}=r_\vir\frac{(1+z)^2}{d_L(z)},
\end{equation}
where we adopt a fitting formula \citep{pen99} for $d_L(z)$.  With
this choice, the only free parameters needed to specify the full gas
density profile, $\rho_\gas(r)$, are $M_\vir$, $\alpha$, and $z$.

Let us now introduce a truncation of the density (and temperature)
profile (\ref{eq:y_dm}) to account for the shock front near the virial
radius of the cluster. As pointed out above (see eq.~\ref{eq:rho_gas0}),
such a truncation was already needed for calculating the normalization of the
gas density.  Let us assume that the density has a linear cutoff
between radii $x_\mx-D$ and $x_\mx+D$. At $x_\mx+D$ the density
attains the background value. Since this value is small compared to
$y_\dm(c)$ for realistic clusters, we set both the gas density and
temperature at $x>x_\mx+D$ to zero.  Since the gas density profile
traces the dark matter profile, the linear density cutoff can be
imposed directly on the gas density.  In what follows, we will refer
to the original density profile (eq.~\ref{eq:y_gas}) as
${y_\gas}_0(x)$. We will denote the density profile with a cutoff by
$y_\gas(x)$, defined by
\begin{equation}
y_\gas(x)= {y_\gas}_0(x) W(x),
\end{equation}
where
\begin{equation}
W(x)= \left\{
\begin{array}{lll}
  1 & \text{if~~} x<x_\mx-D \\
  {[(x_\mx+D)-x]}/{(2D)} & \text{if~~} x_\mx+D<x<x_\mx-D \\
  0 & \text{if~~}x_\mx+D<x \\
\end{array}\right.
\end{equation}

For consistency, the $\rho_\gas(0)$ gas density normalization has to
be recalculated from equation~(\ref{eq:rho_gas0}) using the mass
$m(x)$ enclosed by the density profiles truncated with a given choice
of $x_\mx$ and $D$ (see eq.~\ref{eq:m(x)}). In practice,
$\rho_\gas(0)$ is sensitive to $x_\mx$ only, with the $D$ dependence
nearly negligible. Thus, for calculating the normalization
$\rho_\gas(0)$, we always assume $D=0$ in
equation~(\ref{eq:rho_gas0}).


\section{SZ surface brightness profiles}
\label{sec:szprofiles}

The two--dimensional SZE intensity profile $I(x)$, as a function
of projected radial distance $x$ away from the cluster center, is
often expressed as a small temperature change of the CMB spectrum.
The SZE corresponding temperature increment $\Delta T(x)$ (which
can be negative) is defined by
\begin{equation}\label{eq:I-T}
I(x)=\frac{\partial I_{CMB}}{\partial T_{CMB}}\Delta T(x) .
\end{equation}
Solving for the SZE temperature yields \citep{carl02}
\begin{equation}\label{eq:surfacebrightness0}
\Delta T(x)=f(\theta)y_C(x)T_{CMB} ,
\end{equation}
where $\theta={h\nu}/{k_\mathrm{B}T_{\mathrm{CMB}}}$ is the
dimensionless frequency, $y_C$ is the Compton parameter, and
$f(\theta)$ is given by
\begin{equation}
f(\theta)=\left(\theta\frac{e^\theta+1}{e^\theta-1}-4\right)
(1+\delta_{SZE}(\theta,T_e)) .
\end{equation}
The $\delta_{SZE}(x,T_e)$ term is the relativistic correction to the
frequency dependence, which is negligible at 100 GHz, but becomes
important at higher frequencies ($\gsim 250$ GHz).
The Compton $y_C$-parameter is defined as
\begin{equation}\label{eq:Compton}
y_C(x)=\frac{k_B \sigma_T}{m_e c^2} \int n_e T_e  \D l
\end{equation}
where $\sigma_T$ is the Thompson cross-section, $n_e$ is the electron
number density, $T_e$ is the electron temperature, $k_B$ is the
Boltzmann constant, $m_e c^2$ is the electron rest mass energy, and
the integration is along the line of sight (i.e. along
$r=\sqrt{x^2+l^2}$ for a given $x$). We calculate the electron
temperature, $T_e$, with $T_e=T_\gas$ using
equation~(\ref{eq:mass-temperature}), and the number density is given
by
\begin{equation}\label{eq:n_e}
n_e(x)=\frac{\rho_\gas(0)y_\gas(x)}{\mu m_\mathrm{p}}
\end{equation}
where $m_\mathrm{p}$ is the proton rest mass, and $\mu=0.59$ for an
ionized H-He plasma with $25\%$ Helium abundance by mass.

\begin{figure}
  \centering\mbox{
    \includegraphics[width=7cm]{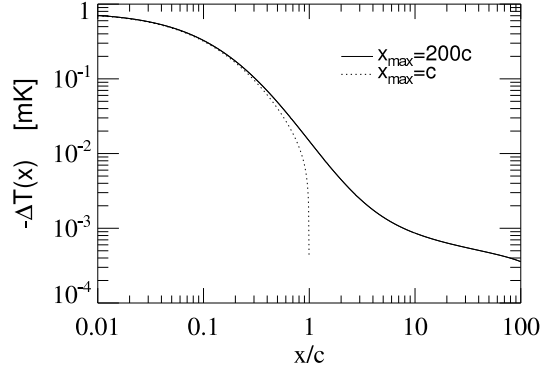}}
  \caption[]{\label{fig:surfacebrightness} Predicted SZ surface
  brightness profiles for $M_\vir=10^{15}M_\odot$, $z=0.1$,
  $\alpha=1.5$, and $\nu=100\Hz$. The solid curve shows the profile
  without a cutoff (the calculation used $x_\mx=200$ and $D=0$ in
  practice), and the dotted line corresponds to the SZ profile with
  $x_\mx=c$ and $D=0$. Equation~(\ref{eq:T(x)}) was evaluated with
  $\Delta T(0)=0.7\mK$.  }
\end{figure}

Substituting (\ref{eq:temperature}), (\ref{eq:mass-temperature}),
and (\ref{eq:n_e})
 in (\ref{eq:Compton}) the SZ surface
brightness profile (\ref{eq:surfacebrightness0}) is separable into
a dimensionless integral with spatial dependence and a constant
coefficient
\begin{equation}\label{eq:T(x)}
\Delta T(x)= \Delta T_s Y(x) ,
\end{equation}
where
\begin{equation}\label{eq:T_s}
\Delta T_s=f(\theta)T_{CMB}\sigma_T \frac{1}{m_e
c^2}\frac{GM_\vir}{3c} \rho_\gas(0)\eta
\end{equation}
and
\begin{equation}\label{eq:Y(x)}
Y(x)= 2\int^{\sqrt{(x_\mx+D)^2-x^2}}_{0} \D l\,
[y_\gas(\sqrt{x^2+l^2})]^\gamma.
\end{equation}

\begin{figure}
\centering\mbox{
\includegraphics[width=7cm]{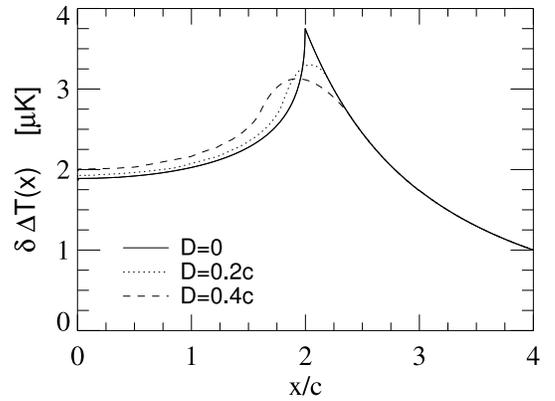}}
\caption[]{\label{fig:surfacebrightnessdifference} The difference
    between the normalized brightness profiles with and without a
    cutoff, for various $D$ values. The difference is taken between
    the $\Delta T(x)$ profiles with different $x_\mx$ and $D$
    parameters, but identical $(M_\vir, \alpha, z)=(10^{15}
    M_\odot,1.5,0.1)$, and $\Delta T_s$ values. The SZ profiles with
    $x_\mx=2c$ and $D=0$, $0.2c$, and $0.4c$ were compared with the
    same profile without a cutoff, for which $x_\mx=200$ and $D=0$ was
    used in practice. The fact that $\Delta T-\Delta T^{\rm fid}$ is
    nonzero at $x=0$ is a consequence of the $x_\mx$ dependence of
    Y(x).}
\end{figure}

Figure~\ref{fig:surfacebrightness} depicts the $\Delta T(x)$
profile for $\alpha=1.5$ with and without a cutoff. Without the
cutoff, the $y_\gas(x)$ and $\Delta T(x)$ profile has a nonzero
limit for large $x$. Various choices of $x_\mx$ can be compared by
analyzing the difference between the associated intensity
profiles. Figure~\ref{fig:surfacebrightnessdifference} shows the
difference between pairs of profiles with $x_\mx=2c$ and without a
cutoff, using either $D=0$, $0.2c$, or $0.4c$ for the truncated
profile. All other parameters, such as $M_\vir$, $\alpha$, $z$,
and $\Delta T_s$ were taken to be equal. Note that the $\Delta
T(x)$ profile is obtained from $Y(x)$ by multiplying by the
$\Delta T_s$ coefficient. According to equation~(\ref{eq:rho_gas0}),
various $x_\mx$ choices generally lead to unequal $\Delta T_s$
values, implying a large nonzero difference between the $\Delta
T(x)$ profiles at $x=0$. In principle, this difference is
physical, reflecting the normalization criterion we chose (namely
that the clusters contain the average baryon fraction within the
virial shock -- no mechanism is known to segregate baryons from
dark matter on these large scales). Since the normalization is
model--dependent, one might worry that it will effect our analysis
below, where we study the detectability of the virial shocks.
However, we find that the information from the normalization does
not dominate our results (see discussion below).

Table \ref{tab:T} shows the central $\Delta T(0)$ values for several
different cluster masses and density profile slopes, and for various
frequencies.  We find that the $\alpha=1.5$ dark matter profile yields
$\propto 1.6\times$ larger $\Delta T(0)$ values than the $\alpha=1$
model.  We can elucidate the source of this increment by tracking the
differences between the two models in equations (\ref{eq:T(x)}) and
(\ref{eq:T_s}) while fixing the other parameters at
$M_\vir=10^{15}M_\odot$, $z=0.3$, $x_\mx=2c$, and $D=0.01$. The
difference is caused by the change in the product $\rho_\gas(0)\eta
Y(0)/c$. First, since the dark matter model with $\alpha=1.5$ is
concentrated more in the central region, the gas density and the gas
temperature is expected to be higher in the center for
$\alpha=1.5$. Indeed, $\rho_\gas(0)=4.5\times 10^{4}\rho_{BG}$ for
$\alpha=1.5$, while it is $1.0\times 10^4\rho_{BG}$ for
$\alpha=1$. Second, the $\eta$ parameter is simply proportional to the
central gas temperature $T_\gas(0)$ (\ref{eq:mass-temperature}), which
is somewhat higher (by 21\%) for $\alpha=1.5$. Third, the $1/c$ factor
further increases the difference by 70\%. Finally, the increment
caused by variables localized to the center is smeared by $Y$, which
accounts for the fact that the observation measures the projection of
the intensity along the line of sight. In particular, $Y(0)=0.13$ for
$\alpha=1.5$ and $Y(0)=0.71$ for $\alpha=1$. Therefore, the resulting
increase in the central SZE temperature is 60\% for $\alpha=1.5$.


\begin{deluxetable}{cccccc}
\tablecolumns{6} \tablewidth{0pt} \tablecaption{\label{tab:T}
Central SZE decrement temperatures} \tablehead{
\colhead{$M_\vir/M_\odot$} & \colhead{ $\alpha$} &
\colhead{$100\,\GHz$} & \colhead{$150\,\GHz$} &
\colhead{$200\,\GHz$} & \colhead{$250\,\GHz$}} \startdata
      $10^{13}$ & 1     & -0.019 & -0.012 & -0.003 & 0.006 \\
      $10^{13}$ & 1.5   & -0.029 & -0.018 & -0.005 & 0.010 \\
      $10^{14}$ & 1     & -0.106 & -0.067 & -0.018 & 0.036 \\
      $10^{14}$ & 1.5   & -0.173 & -0.110 & -0.030 & 0.059 \\
      $10^{15}$ & 1     & -0.609  & -0.385 & -0.106 & 0.207 \\
      $10^{15}$ & 1.5   & -1.017  & -0.643 & -0.178 & 0.345 \\
\enddata
\tablecomments{The SZE decrement temperatures are given in mK. We
assumed $x_\mx=2c$, $D=0.01c$ and $z=0.3$. $\Delta T(0)$ varies by
less than $20\%$ for the range $c<x_\mx<10c$.}
\end{deluxetable}

\section{Noise Estimates for Future SZ Brightness Maps  }
\label{sec:noise}

We shall now summarize the primary observational difficulties for
detecting virial shocks.\footnote{This discussion ignores obvious
additional theoretical modelling difficulties, such as departures
from the spherical cluster models adopted here. These will be
discussed in \S~\ref{subsec:varyparameters} and
\S~\ref{sec:conclusions} below.} The possible sources of
contamination are detector noise, atmospheric fluctuations, and
other uncertainties of astrophysical origin. Inevitable
astrophysical contaminants include radio point sources, bright
foreground (and background) clusters along the line of sight, and
the primary and secondary CMB anisotropies.

As pointed out in \S~\ref{sec:alma} above, the atmospheric and
detector uncertainties will not be crucial at moderate frequencies, as
the ALMA system can reduce its noise level below $10\muK$ for a $2"$
beam at $100\GHz$ in an integration time of $\sim 20\rm hr$.

Similarly, bright extragalactic foregrounds will not pose a great
barrier against the observation of virial shocks. The radio
point sources cover only a negligible portion of the set of relevant
pixels, and their removal reduces the efficiency of a high resolution
detection by a negligible amount. The flux of the brightest clusters
identified in separate wide--field X-ray and SZ surveys can also be
subtracted from the image. Moreover, uncertainties from the
imprecision of this information can be minimized by choosing galaxy
clusters with no obvious overlapping foreground clusters.

The magnitude of uncertainties from primary and secondary CMB
anisotropies can be assessed by the analysis of the corresponding
angular power spectra. The primary CMB anisotropy supplies the
dominant power on large-scales, but is exponentially suppressed well
below $10\muK$ for scales under $\sim 10'$. The angular radius of the
clusters decreases below this scale at $z>0.19$ for
$M_\vir=10^{15}M_\odot$ and $z>0.08$ for $10^{14} M_\odot$ (eq.
\ref{eq:Theta}). We shall restrict our calculations to larger
redshifts, where secondary CMB anisotropies will likely dominate.  The
two largest sources of the secondary CMB anisotropies are the
fluctuations from unresolved background clusters (i.e. overlapping
faint clusters of varying temperatures, sizes, and numbers along the
line of sight) and the Ostriker-Vishniac (OV) effect. The OV
contribution can be approximated by a flat band power contribution of
$1 \mu K$ \citep{HD02}. According to both theoretical and numerical
studies \citep{hold01,hold02,KS02,spring01}, the rms power of the
unresolved thermal SZE contamination is between $1$ and $6 \mu K$ on
arcminute scales. For low--mass clusters this noise background is
strongly non-Gaussian, and there is a significant amount of
uncertainty in the small-scale power spectrum, due to
non-gravitational effects such as gas cooling and feedback from star
formation. For a conservative estimate, we shall assume that the
fluctuations around the SZ profile are given by white Gaussian noise
with amplitude $\sigma_N=10\muK$.


\section{Significance of Detecting Virial Shocks}
\label{sec:edgedetection}

In this section, we consider the detectability of virial shocks in
galaxy clusters.
The basic input to this analysis is a (mock) SZ surface brightness
map of the cluster. We compute the SZ profile $\Delta T^{\rm
fid}(x)$ according to equation~(\ref{eq:T(x)}) in a mock
``fiducial'' model, in which the cluster is assumed to be
spherically symmetric with a well--defined edge, at the position
of the virial shock.  This fiducial
model is uniquely described by five parameters, specifying the
global properties of the cluster $(M_\vir,z,\alpha)$ and the
location and sharpness of the virial shock $(x_\mx^{\rm fid},D)$.
For concreteness, in the numerical calculations, we use a single,
fixed measurement frequency of $100\,\GHz$.  This value was chosen
to minimize the integration time needed for ALMA, while keeping
the total SZ decrement around its maximum in the nonrelativistic
regime.

Our basic task is to evaluate the significance, given the $\sim10\mu$K
noise, at which $\Delta T^{\rm fid}(x)$ can be distinguished from a
different model profile $\Delta T^{\rm test}(x)$, computed in a test
model that does not include a shock. This task immediately raises the
question: what should the no--edge benchmark model be for this
comparison?

The simplest approach would be to choose the underlying density
profile to be identical to that in the fiducial model within
$x\lsim x_\mx^{\rm fid}$.  In the absence of a virial shock,
however, the smooth density profile would extend to some large
radius. This could be expressed by choosing the same parameters
$(M_\vir,z,\alpha,D)$ as in the fiducial model, but replacing the
location of the virial shock by $x_{\mx}^{\rm
test}\rightarrow\infty$. However, a complication is that our
procedure to self--consistently normalize the density profile
(eq.\ref{eq:rho_gas0}) would then fail, due to the divergence of
the mass integral (eq. \ref{eq:m(x)}).  To resolve this ambiguity,
we shall adopt either of the following two approaches to define a
smooth, no--edge ``test'' model (hereafter referred to as Model I
and Model II, respectively):

\begin{enumerate}

\item The central gas density $\rho_{\rm gas}(0)$ is chosen to
equal the central density in the fiducial model. Equivalently, the
central decrements $\Delta T_s$ are assumed to be the same for the
two hypotheses.  The value of $x_{\mx}^{\rm test}$ is then
irrelevant, as long as it is chosen to be sufficiently large,
$x_{\mx}^{\rm test}\gg c$ (in practice, we find that $x_{\mx}^{\rm
test}>10c$ suffices).

\item The location of the virial shock $x_{\mx}^{\rm test}$ is
chosen to equal the radius at which the cluster gas density equals
the background baryon density. This value is between $4.5\leq
x_{\mx}^{\rm test}/c\leq 6$ for typical clusters with virial mass
$10^{13}M_\odot\leq M_\vir\leq 10^{15}M_\odot$, at any redshift
and $\alpha$ (either 1 or 1.5).  The gas density normalization is
calculated self--consistently, using equation~(\ref{eq:rho_gas0}).
The profile outside $x_{\mx}^{\rm test}$ is truncated with $D=0$.
Note that the temperature decrement scale $\Delta T_s$ and the
central temperature decrement $\Delta T(0)$ of this model are both
different from the fiducial model (by $\sim 10\%$ for $\Delta
T(0)$ for $x_{\mx}^{\rm fid}=c$).

\end{enumerate}

Model II has the advantage of a closed and self--consistent
theoretical description, where $\rho_\gas(0)$ and thus $\Delta
T_s$ is derived directly from a set of underlying assumptions.
However, we note that the SZ brightness profile in this model
differs from the fiducial model because of the ``renormalized''
value of the temperature decrement $\Delta T_s$.  Hence, the
distinguishability of test model II from the fiducial model will
have an additional indirect dependence on the presence of the
virial shock, through the effect of the virial shock on the
overall normalization of the density profile. This inference will
then be model--dependent (i.e. on our particular assumption about
the relationship between the location of the virial shock and the
normalization of the gas density), and also subject to
uncertainties due to physical effects (i.e. preheating;
\citealt{hold01}) that can significantly modify the inner density
profile.  Nevertheless, we find in our calculations below that the
change in the temperature due to its renormalization is relatively
small, and changes in the SZ brightness profile are only
significant in the inner regions.  Overall, the temperature change
adds nearly negligibly to the detectability of the virial shock.
Hence, in practice, model II is a useful alternative to model I.
The latter, by construction, isolates the effect on the
surface brightness profile of a sharp discontinuity itself.

The expected SZ profile of a cluster with or without a cutoff can
be calculated from equation~(\ref{eq:T(x)}) in the fiducial model, and
compared to the profile in either test model.  In the next two
subsections, we quantify the significance for detecting the
difference between these pairs of models.

In these estimates, we will assume that the interferometer has fully
synthesized an aperture, yielding a narrow ($\sim 2''$) effective
point response function (PRF).  As discussed above, this assumption is
realistic for $\sim 20$hr interferometric observations with
ALMA. Thus, we will employ the simple, single--dish technique of
estimating $S/N$, in which the image of the extended source can be
thought of as being composed of $K$ independent pixels (where each
pixel covers a solid angle of $\sim 12$ square arcseconds).

Finally, as stated above, in the simplest estimates, we fix the
parameters $(M_\vir,z,\alpha,\Delta T_s)$ in Model I, and the
parameters $(M_\vir,z,\alpha)$ in Model II, to be the same as in
the fiducial model.  While $M_\vir$ and $z$ can be independently
estimated from gravitational lensing, X--ray, and optical data,
the decrement normalization $\Delta T_s$ of the fiducial model
will not be known a--priori.  Therefore, to obtain a more
realistic estimate for the detectability of the virial shock, in
section \ref{subsec:varyparameters} below, we will relax these
assumptions. More specifically, we will obtain fits to the mock SZ
profile of the fiducial model, by allowing $\Delta T_s$ (as well
as the concentration parameter $c$; see further discussion below)
to be free parameters in test model I.

\subsection{Significance of Shock Detection in Fixed Test Model}
\label{subsec:fixedmodel}

We shall first derive a simple estimate for the significance for
choosing between profiles with and without a cutoff, using the
comparison between test model I and the fiducial model.  Since all
of the parameters of the test model are fixed, this exercise will
yield an estimate for the significance of the change in the SZ
surface brightness maps caused by the presence of a shock front.
The presence of the shock, in general, is more difficult to infer
when the test model's parameters are not assumed to be known
\textit{a priori}. This question will be taken up in
\S~\ref{subsec:varyparameters} below.

The cutoff radii in the fiducial and test models are taken here to be
$x_{\mx}^{\rm fid}=c$ and $x_{\mx}^{\rm test}=10c$, respectively
(although our results below do not depend on the latter choice).  The
effective ``signal to noise ratio'' is calculated by comparing the
difference in the ``signal'' $\delta \Delta T(x)=\Delta T^{\rm
test}(x)-\Delta T^{\rm fid}(x)$ to the noise power in the regular
two-dimensional angular space.

The effective signal to noise ratio corresponding to a single
pixel at the projected radius $x$ is
\begin{align}\label{eq:S/N(K=1)}
\left(\frac{S}{N}\right)^2_{K=1}&=\frac{\max_x(\Delta
T^{\rm test}(x)-T^{\rm fid}(x))^2}{\sigma_N^2}
\\
&=\left(\frac{\Delta T^{\rm fid}(0)}{\sigma_N}\right)^2\frac{\max_x(Y^{\rm test}(x)-Y^{\rm fid}(x))^2}{Y^{\rm fid}(0)^2}\\
&=\left(\frac{\Delta T(0)}{\sigma_N}\right)^2\frac{(Y^{\rm test}(x_\mx^{\rm fid})-Y^{\rm fid}(x_\mx^{\rm fid}))^2}{Y(0)^2}.
\end{align}
In the last two lines, we have dropped the superscript from $\Delta
T(0)$ and $Y(0)$ because these have the same values in the fiducial
and test models (although $\Delta T(0)$ and $Y(0)$ do depend on
$x_\mx$, the difference between the fiducial and test models is
negligibly small).  We have also adopted $D^{\rm fid}=0$ for the fiducial model,
in which case the maximum of $\delta \Delta T(x)$ occurs at
$x_\mx^{\rm fid}$ (see Fig.~\ref{fig:surfacebrightnessdifference}).
For a massive cluster with $M_\vir=10^{15}M_\odot$, $\alpha=1.5$ and
$z=0.3$, we find $\Delta T(0)=1.06 \mK$, and
\begin{align}
\frac{(Y^{\rm test}(x_\mx)-Y^{\rm fid}(x_\mx))^2}{Y(0)^2}&= 6.0\times 10^{-4}\\
\Delta T(0)^2/\sigma_N^2=1.13\times 10^4,
\end{align}
implying that $S/N\approx 2.6$ for the single best pixel on the
cluster surface, i.e. at the pixel corresponding to the edge of
the cluster at the virial shock.  These values increase somewhat
with increasing $z$.

At an effective angular resolution of $\Delta\phi$, the total number
of resolved pixels covered by the cluster is given by the (solid angle
extended by the circular disk within the shock radius) divided by the
(solid angle extended by a single pixel):
\begin{equation}
K_{\rm tot}=\frac{x_\mx^2}{c^2}\frac{\Theta_\vir^2}{\Delta\phi^2}.
\label{eq:pixelno}
\end{equation}
Note that for this estimate, we consider the cluster to extend out
to $x_\mx^{\rm fid}$; this will result in an underestimate, since
the difference between the fiducial and test SZ profiles extends
out to larger radii (see
Figure~\ref{fig:surfacebrightnessdifference}). Since $\Theta=8.1'$
for $z=0.3$ for a massive cluster, we expect $K_{\rm
tot}=5.9\times 10^4$ pixels for the angular resolution
$\Delta\phi=2"$ of ALMA. Of these $K_{\rm tot}$ pixels, $K_{\rm
edge}\sim (2x_\mx\Theta_\vir)/(c\Delta\phi)$ lie along the
1--pixel wide circular annulus along the virial shock.  A quick
estimate for the total S/N distinguishing the two models is given
by adding the maximal single--pixel S/N in quadrature, i.e.
\begin{equation}\label{eq:S/N heuristic}
\frac{S}{N} \approx \sqrt{K_{\rm edge}}\left(\frac{S}{N}\right)_{K=1}.
\end{equation}

For a massive cluster $M_\vir=10^{15}M_\odot$ at $z=0.3$, we find
$K_{\rm edge}\approx 490$, and the detection of the virial shock
at $x_\mx=c$ with ALMA has a signal to noise ratio $S/N\approx
57$. Since this value is from a single pixel--wide ring around the
virial shock, it is strictly only a lower limit.
Figure~\ref{fig:surfacebrightnessdifference} shows that there is
an annulus of finite $\Delta x_{\rm edge}$ width around the virial
shock with a significant $\delta \Delta T$.  $\Delta x_{\rm edge}$
can be estimated with the radial distance where $\delta\Delta T$
drops to half of its maximum value. A simple formula found from
approximating $\Delta T(x)$ to second order in $x$ is $\Delta
x_{\rm edge} = x^{\rm fid}_\mx/4$, leading to $K_{\rm edge}\sim
({x^{\rm fid}_\mx}\Theta_\vir)^2/(2c^2\Delta\phi^2)$. For a
massive cluster, we obtain $K_{\rm edge}=3.0\times10^4$ and
$S/N=440$.  We find little dependence of this estimate on $\alpha$
for choices of $\alpha=1$ vs. $\alpha=1.5$.  On the other hand, we
find a redshift dependence: larger $z$ decreases $\Theta_\vir$ but
increases $\delta\Delta T(x_\mx^{\rm fid})$. The combined result
is a minimum in the total $S/N$ at $z=0.28$.  Lower--mass clusters
have a smaller angular size and a lower SZ decrement, so $S/N$
decreases with $M_\vir$.  Overall, the virial shock is most
visible for clusters that are either close--by ($z\sim 0$), or at
high redshift ($z\gg 0.28$), and that have a large $M_\vir$ and
$\alpha$.

To improve the simple estimates above, we can average the $S/N$ over
the face of the cluster. This leads to the following approximation for
the average single--pixel $S/N$:
\begin{equation}\label{eq:S/N(K=1) average}
\left(\frac{S}{N}\right)^2_{K=1}= \left(\frac{\Delta
T(0)}{\sigma_N}\right)^2 \frac{\int_0^x \D x'\,2\pi x' (Y^{\rm
test}(x')-Y^{\rm fid}(x'))^2}{\pi x^2Y(0)^2} .
\end{equation}

In Appendix A, we give a rigorous derivation of the $S/N$ for
distinguishing between models in the presence of a Gaussian noise.  We
find that equation~(\ref{eq:S/N(K=1) average}), with the average taken
in the interval $0\leq x \leq x_\mx^{\rm test}$, multiplied by the
square--root of the number of pixels, $\sqrt{K_{\rm tot}}$
(eq.~\ref{eq:pixelno}), indeed gives the formally correct answer.
Accordingly, the significance of distinguishing test model I or
model II from the fiducial model is given by
\begin{equation}\label{eq:S/N optimal}
\frac{S}{N} =  \frac{\sqrt{K}}{\sqrt{A}\, \sigma_N}\times\left\{
\begin{array}{lll}
\Delta T_s \sqrt{\int\D ^2x\, [Y^{\rm test}(x)-Y^{\rm fid}(x)]^2}  & \text{(I)} \\
 \sqrt{\int\D ^2x\, [\Delta T_{s}^{\rm test} Y^{\rm test}(x)-\Delta T_{s}^{\rm fid}Y^{\rm fid}(x)]^2}  & \text{(II)}  \\
\end{array}\right.
\end{equation}
Here $K=(x^{\rm test}_\mx\Theta_\vir)^2/(c\Delta\phi)^2$ is the
number of pixels, and $A=\pi {x^{\rm test}_\mx}^2$ is the area
covering the cluster's surface.
Note that the only difference between Models I and II is that in
the latter case, the temperature scale $\Delta T_{s}^{\rm test}$
is recalculated using the cutoff location $x_{\mx}^{\rm test}$.
The right--hand--side of equation~(\ref{eq:S/N optimal}) can be
compared with Figure \ref{fig:surfacebrightnessdifference}, which
plots the integrand for Model I.

We evaluated the significance given by equation~(\ref{eq:S/N
optimal}) for various choices of fiducial models.  Recall that the
fiducial model is specified by five parameters
$(M_\vir,z,\alpha,x_\mx,D)$. The results are listed in
Table~\ref{tab:S/N optimal1} for three different masses ($M_{\rm
vir}=10^{13},10^{14},10^{15}{\rm M_\odot}$) and two different
inner slopes ($\alpha=1,1.5$).  The other three parameters in this
Table are fixed at the values of $(z,x_\mx,D)=(0.3, c, 0.01c)$.
Figure~\ref{fig:S/N(xmax,D,z)} shows how the $S/N$ varies with $z$
and $x_\mx$ for these masses and slopes (we fixed $D=0.01c$ in
this Figure).

In order to study the sensitivity of the S/N to the choice of the
fiducial parameters of the virial shock itself, in Table~\ref{tab:S/N
optimal2} we repeat each calculation in Table~\ref{tab:S/N optimal1},
except we replace $(x_\mx^{\rm fid},D^{\rm fid})=(c,0.01c)$ by
$(x_\mx^{\rm fid},D^{\rm fid})=(2c,0.5c)$.  The latter choice should
also be regarded as more realistic, given that the strong virial
shocks are found in simulations to be located at $\sim$ twice the
virial radius (e.g. Figure 16 in Bryan \& Norman 1998), and may not be
perfectly sharp.  As Table~\ref{tab:S/N optimal2} shows, the larger
value of $x_\mx^{\rm fid}$ decreases the S/N by about a factor of
two (see Figure~\ref{fig:S/N(xmax,D,z)}).

The main conclusion that can be drawn from Table~\ref{tab:S/N
optimal1} and Figure~\ref{fig:S/N(xmax,D,z)} is that the $S/N$ is
a strong function of the size of the cluster: the virial shock of
a massive cluster with $M_\vir\sim 10^{15}{\rm M_\odot}$ is
detectable at high significance ($S/N~\sim 500$), but the
detection becomes marginal for low--mass clusters or groups with
$M_\vir\sim 10^{14}{\rm M_\odot}$, and clearly impossible for even
smaller objects.

\begin{deluxetable}{ccccccccc}
\tablecolumns{9} \tablewidth{0pt} \tablecaption{Typical $S/N$
ratios for detecting the virial shock \label{tab:S/N optimal1}}
\tablehead{\colhead{$M_\vir$} & \colhead{$\alpha$} &
\colhead{$\Theta$} & \colhead{c} & \colhead{$|\Delta T_{s}^{\rm
fid}|$} & \colhead{$|\Delta T_{s}^{\rm test}|$} & \colhead{$x^{\rm
test}_\mx$} & \colhead{$S/N$} & \colhead{$S/N$} \\
\colhead{$[M_\odot]$} & \colhead{}& \colhead{$[']$} & \colhead{} &
\colhead{$[\mK]$} & \colhead{$[\mK]$} & \colhead{$[c]$} &
\colhead{I} & \colhead{II}} \startdata
$10^{13}$ &  1  & 1.8 & 7.9 & 0.02& 0.02& 4.6 & 0.6 & 0.7\\
$10^{13}$ & 1.5 & 1.8 & 4.6 & 0.2 & 0.2 & 4.7 & 0.6 & 0.8\\
$10^{14}$ & 1   & 3.8 & 5   & 0.1 & 0.1 & 5.2 & 18  & 17\\
$10^{14}$ & 1.5 & 3.8 & 2.9 & 1.3 & 1.1 & 5.4 & 19  & 19\\
$10^{15}$ & 1   & 8.1 & 3.1 & 0.9 & 0.8 & 5.8 & 550 & 440\\
$10^{15}$ & 1.5 & 8.1 & 1.8 & 8.3 & 6.9 & 6   & 570 & 480 \\
\enddata
\tablecomments{In all cases, we assume $z=0.3$, $x^{\rm
fid}_\mx=c$ and $D^{\rm fid}=0.01c$.  The self--consistently
derived $x^{\rm test}_\mx$ value is used for Model II, whereas a
fixed $x^{\rm test}_\mx=10c$ was assumed for Model I.}
\end{deluxetable}

\begin{deluxetable}{ccccccccc}
\tablecolumns{9} \tablewidth{0pt} \tablecaption{Typical $S/N$
ratios for detecting the virial shock \label{tab:S/N optimal2}}
\tablehead{\colhead{$M_\vir$} & \colhead{$\alpha$} &
\colhead{$\Theta$} & \colhead{c} & \colhead{$|\Delta T_{s}^{\rm
fid}|$} & \colhead{$|\Delta T_{s}^{\rm test}|$} & \colhead{$x^{\rm
test}_\mx$} & \colhead{$S/N$} & \colhead{$S/N$} \\
\colhead{$[M_\odot]$} & \colhead{}& \colhead{$[']$} & \colhead{} &
\colhead{$[\mK]$} & \colhead{$[\mK]$} & \colhead{$[c]$} &
\colhead{I} & \colhead{II}} \startdata
$10^{13}$ &  1  & 1.8 & 7.9 & 0.02& 0.02& 4.5 & 0.3 & 0.4\\
$10^{13}$ & 1.5 & 1.8 & 4.6 & 0.2 & 0.2 & 4.6 & 0.3 & 0.4\\
$10^{14}$ & 1   & 3.8 & 5   & 0.1 & 0.1 & 5.1 & 8.9 & 10\\
$10^{14}$ & 1.5 & 3.8 & 2.9 & 1.2 & 1.1 & 5.3 & 9.6  & 12\\
$10^{15}$ & 1   & 8.1 & 3.1 & 0.9 & 0.8 & 5.7 & 280 & 240\\
$10^{15}$ & 1.5 & 8.1 & 1.8 & 7.9 & 7.0 & 5.9 & 300 & 280 \\
\enddata
\tablecomments{Same as Table~\ref{tab:S/N optimal1}, except we
assume $x^{\rm fid}_\mx=2c$ and $D^{\rm fid}=0.5c$.}
\end{deluxetable}

\begin{figure}
\centering\mbox{
\includegraphics[width=7cm]{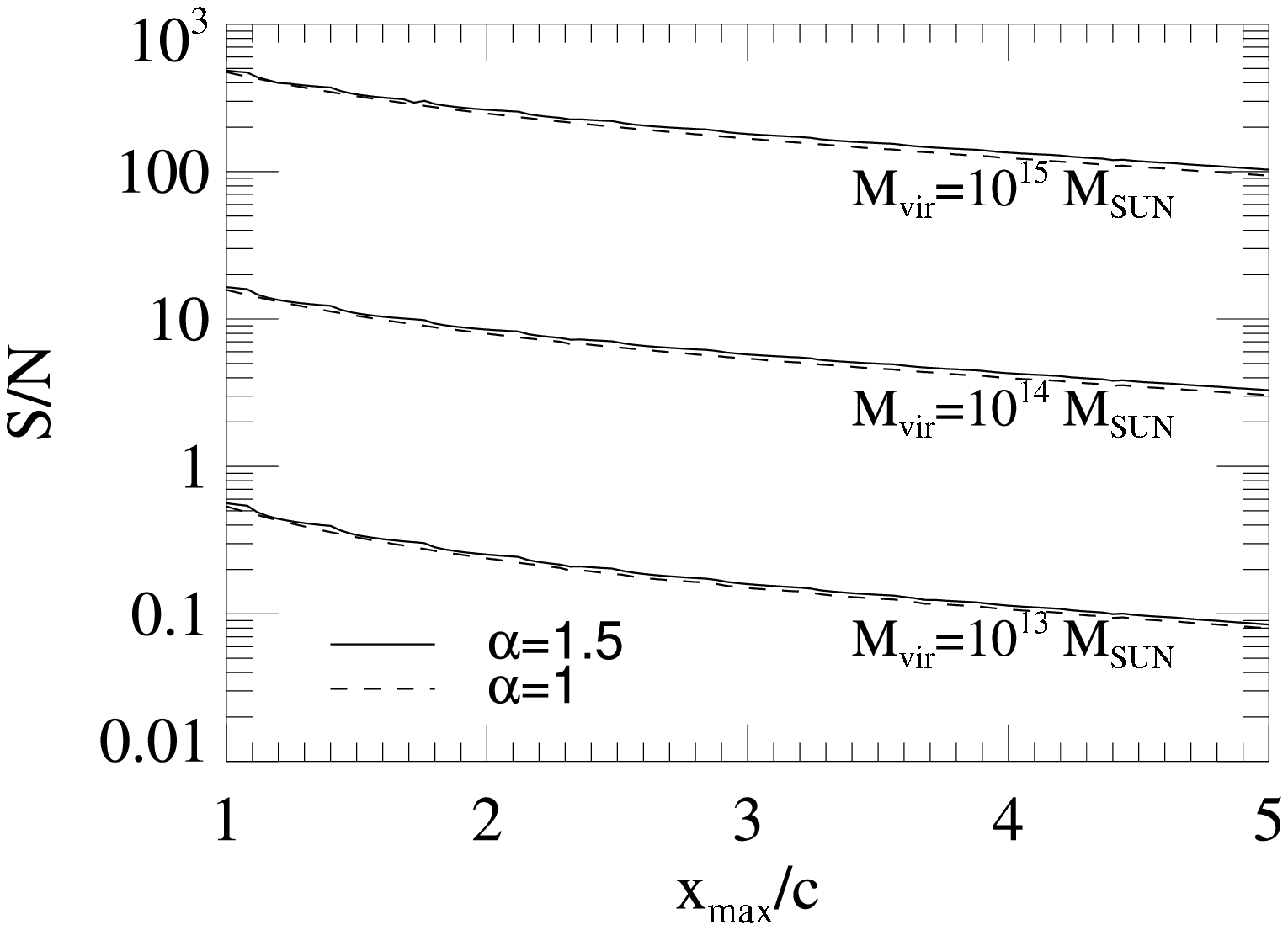}}\\
\centering\mbox{
\includegraphics[width=7cm]{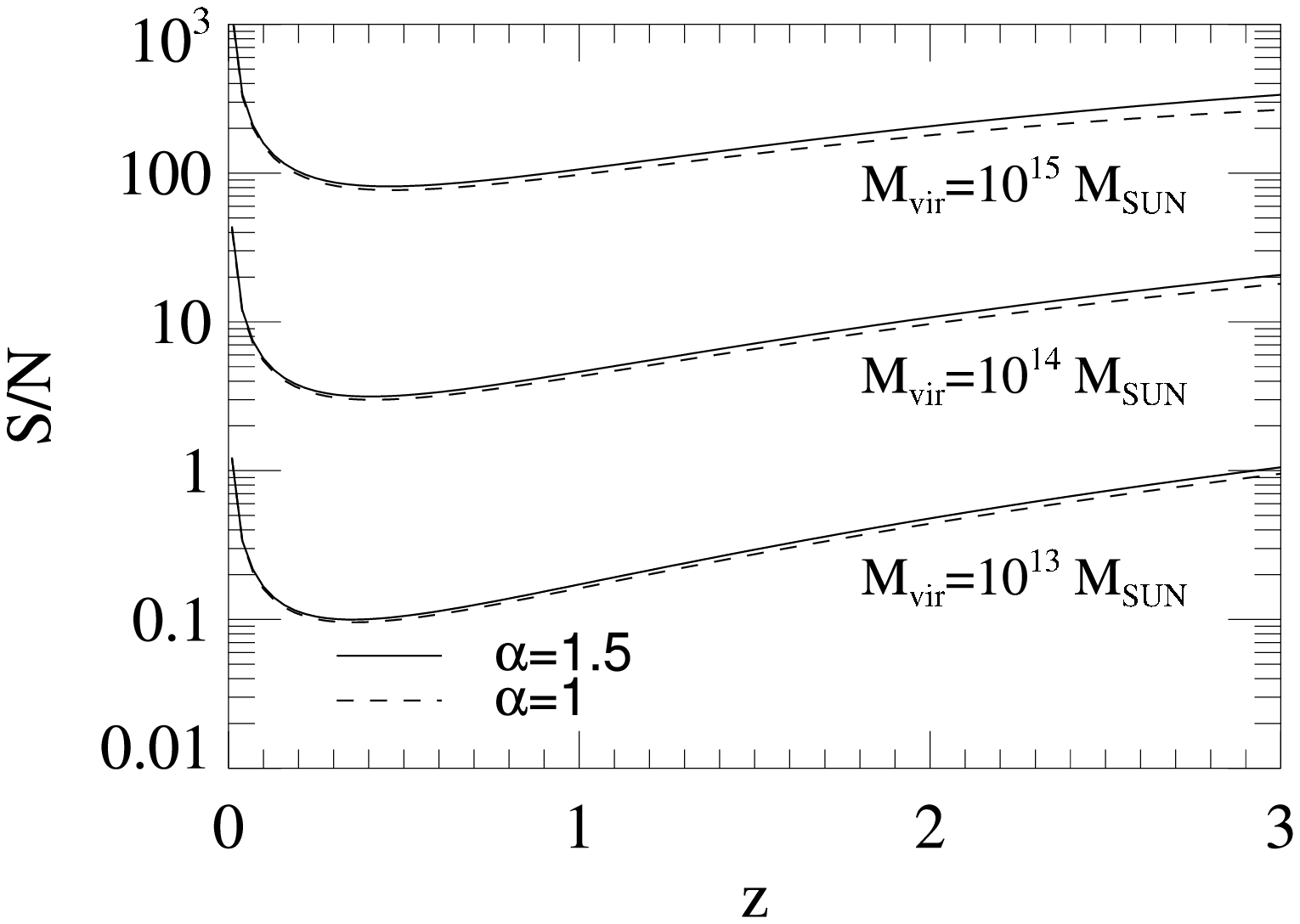}}
\caption[]{\label{fig:S/N(xmax,D,z)}
    The signal to noise ratio for various cluster masses
    as a function of $x_\mx$ \textit{(upper panel)} and
    $z$ \textit{(lower panel)}.
    Each curve has only one parameter changing, the other
    parameters are fixed at the values listed in
    Table \ref{tab:S/N optimal1}.}
\end{figure}

We also find that the $S/N$ is nearly identical for different
choices of $\alpha$ and $D$. However,
Figure~\ref{fig:S/N(xmax,D,z)} shows that $S/N$ decreases with
$x_\mx$, and it is a non--monotonic function of redshift with a
minimum at $z\sim 0.3$. Two of our findings are somewhat
counterintuitive: cluster edges are equally visible for smeared
edges, and are more visible for more distant clusters, even though
these have a smaller angular size.  The first of these results can
be explained by recalling the definition of the cutoff. Increasing
$D$ does not change the total cluster SZ intensity. If the cutoff
shape is known exactly prior to the observation, as it is assumed
in this section, increasing $D$ (while fixing $\Delta T_s$) does
not come closer on average to a profile without a cutoff.

\begin{figure}
\centering\mbox{
\includegraphics[width=7cm]{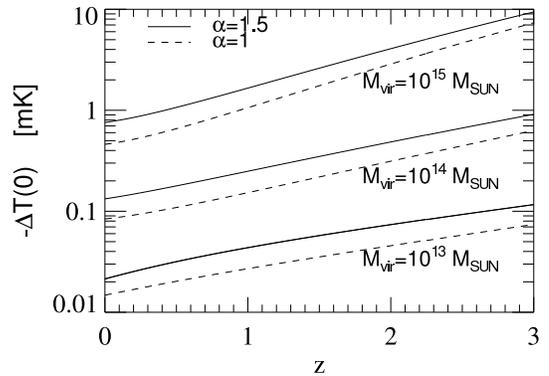}}\\
\caption[]{\label{fig:z}
    The central SZ decrement $\Delta T(0)$ for
    various cluster masses ($M_{\rm vir}$) and power--law slopes ($\alpha$)
    of their inner density profiles,  as a function of redshift.
    Each curve has only one parameter changing, the other fixed
    parameters correspond to those listed in
    Table \ref{tab:S/N optimal1}.
    }
\end{figure}

To explain the second peculiarity, the behavior of $S/N$ as a
function of $z$, we considered the $z$--dependence of each
individual factor that determines the $S/N$.  The most puzzling
feature, i.e. the increase of $S/N$ towards high redshifts at
$z>0.3$, is attributable to an increase in the temperature scale
and in the central temperature decrement $\Delta T(0)$ (shown in
Figure~\ref{fig:z}). Although the angular radius $\Theta$
decreases somewhat towards high--$z$, the temperature increase
dominates. As a result, more distant clusters at $z>0.3$ have a
higher contrast, which makes the virial shock more visible even
though the clusters have a smaller angular size.\footnote{ For
still larger redshifts, when then cluster's angular size drops
below the angular resolution of the instrument, the $S/N$ will
have a cutoff. However, at the $\Delta\phi=2"$ resolution of ALMA,
this critical redshift is well beyond the epoch when clusters form
in a sensible cosmology.}  At low--redshift ($z\lsim 0.3$), the
situation is reversed, and the strong increase towards $z=0$ in
the angular size is the dominant effect, making the virial shock
of very nearby clusters more detectable.

Table~\ref{tab:S/N optimal1} also compares the $S/N$ results for
Model I and II.  The two models lead to approximately equal $S/N$
detection ratios. This is somewhat surprising, since the
renormalization in the temperature scale $\Delta T_s$ and the
difference in the profile shape $Y(x)$ in Model II both introduce
additional differences from the surface brightness $\Delta T(x)$
in the fiducial model. We find that these additional differences
can be significant, as high as $2.5\sigma$ for each pixel near the
central region $x\sim 0$. However, the difference is only at the
level of $\sim 0.15 \sigma$ per pixel at the radius $x\sim x_\mx$
that dominates the cumulative $S/N$.

In conclusion, the estimates in this section indicate that the
temperature decrement difference across a virial shock front
causes a difference of approximately $\Delta T \sim 10\mu$K in the
SZ surface brightness maps. This is a small difference,
approximately at the level of the expected noise.  Hence, in
single neighboring pixels, the shock fronts will not stand out to
be detectable. However, shock fronts are coherent structures
extending over many pixels; in the simple spherical models
considered here, the cumulative $S/N$ ratio is sufficiently high
to infer the presence of the shock front with a high--resolution
instrument, such as ALMA.  The presence of shocks in medium--sized
to massive clusters should be clearly inferable, as long as the
shocks are located near the virial radius. If the density cutoff
is farther in the outer region, the virial shock will only be
observed for relatively massive clusters, with $M>{\rm few}\times
10^{14}M_\vir$. The detection significance is nearly independent
of $\alpha$ and $D$.  The virial shocks of clusters at $z\sim 0.3$
are least detectable, with nearly an order of magnitude increase
in the $S/N$ for $z\approx 0$ or $z\approx 3$ compared to the
minimum value at $z=0.3$ Assuming that the noise is uncorrelated
and Gaussian for different frequencies, measuring at many
different frequencies the signal to noise ratio can be increased
by the square--root of the number of available independent
frequency channels.

\subsection{Significance of Shock Detection in Variable Test Model}
\label{subsec:varyparameters}

The main deficiency of the signal--to--noise ratio analysis in the
previous subsection is that at least some of the parameters of the
test model will not be known a priori.  This, in general, will
make the cluster's virial shock more difficult to detect, since
one may be able to adjust the variable parameters of the no--edge
test model to better mimic the fiducial model that includes the
shock. The mass $M_\vir$ and redshift $z$ of the cluster can be
independently estimated (e.g. from gravitational lensing, X--ray,
and optical data), and the inner density profile slope $\alpha$
has a relatively small impact on our conclusions.  Hence, we still
fix the values of these parameters in this section.  Although the
SZ temperature decrement scale $\Delta T^{\rm test}_s$ can, in
principle, be derived in the fiducial model, it will not be known
a--priori. We therefore allow it to be a free parameter. In
addition, the concentration parameter of the cluster $c$, which,
in principle, can be ultimately derived from simulations, is still
somewhat uncertain. Therefore, we allow $c^{\rm test}$ to be the
second of our two free parameters of the test model in this
section. The maximum radius used for the comparison of the SZ
profiles including and lacking a cutoff is $x^{\rm test}_\mx$, the
value used in Model II previously. Recall that this is the radius
at which the baryon fraction equals the global value $\Omega_{\rm
b}/\Omega_\dm$.

A different complication is that our analysis relies on mock data,
rather than actual data.  Therefore, in the presence of noise, our
``data'' should not be a fixed set of numbers (i.e. the profile
$\Delta T(x)$ in the fiducial model). Rather, it must be a (set
of) probabilistic variables that have a distribution and finite
width.

A rigorous signal--to--noise analysis that incorporates these two
complications is presented in Appendix B.  In this case, the $S/N$
of distinguishing the test model from the fiducial model is itself
a random variable (rather than a fixed number, as we have hitherto
assumed).   However, in Appendix B we show that the expectation
value of $S/N$ obeys the following bounds:
\begin{equation}\label{eq:S/N bounds}
\frac{4}{5}\frac{S_0^2}{\sigma_N^2} \leq \frac{S_{\rm exp}^2}{N^2}
\leq \frac{4}{5}\frac{S_0^2}{\sigma_N^2} + 4.8 ,
\end{equation}
where $S_0^2$ is defined by
\begin{equation}
S_0^2=\frac{K}{A\,\sigma_N^2}\min_{p^{\rm test}} \int \D^2x\,
    [\Delta T^{\rm test}(x,p^{\rm test}) - \Delta T^{\rm fid}(x,p^{\rm fid})]^2.
\end{equation}
The term $p^{\rm test}$ refers to the two variable parameters of
the test model (i.e. $\Delta T^{\rm test}_s$ and $c^{\rm test}$),
and $p^{\rm fid}$ refers to the fiducial values $\Delta T^{\rm
fid}_s$, $c^{\rm fid}$, $x^{\rm fid}_\mx$, and $D^{\rm fid}$. This
distinction is necessary since the corresponding parameters
need not be equal. For a conservative estimate we shall use the lower
bound in equation (\ref{eq:S/N bounds}). The expected value of the
signal to noise ratio is therefore estimated by
\begin{equation}\label{eq:signal to noise advanced}
\frac{S_{\rm exp}}{N} =
\frac{\sqrt{4K}}{\sqrt{5A}\,\sigma_N}\min_{p^{\rm test}}
\sqrt{\int \D^2x\,
    [\Delta T^{\rm test}(x,p^{\rm test}) - \Delta T^{\rm fid}(x,p^{\rm fid})]^2}.
\end{equation}
This should be compared with equation~(\ref{eq:S/N optimal})
describing the signal to noise ratio for fixed parameter values.
The differences are the overall $\sqrt{4/5}$ factor and taking the
minimum over the $p^{\rm test}$ parameter space. We shall denote
the best fitting test parameters, i.e. at which the integral is
minimal, by $p^{\rm test}_0$.

Equation~(\ref{eq:signal to noise advanced}) measures the
probability that the virial shock of the cluster is observable
even in the face of two additionally variable parameters.

\begin{figure}
\centering\mbox{
\includegraphics[width=7cm]{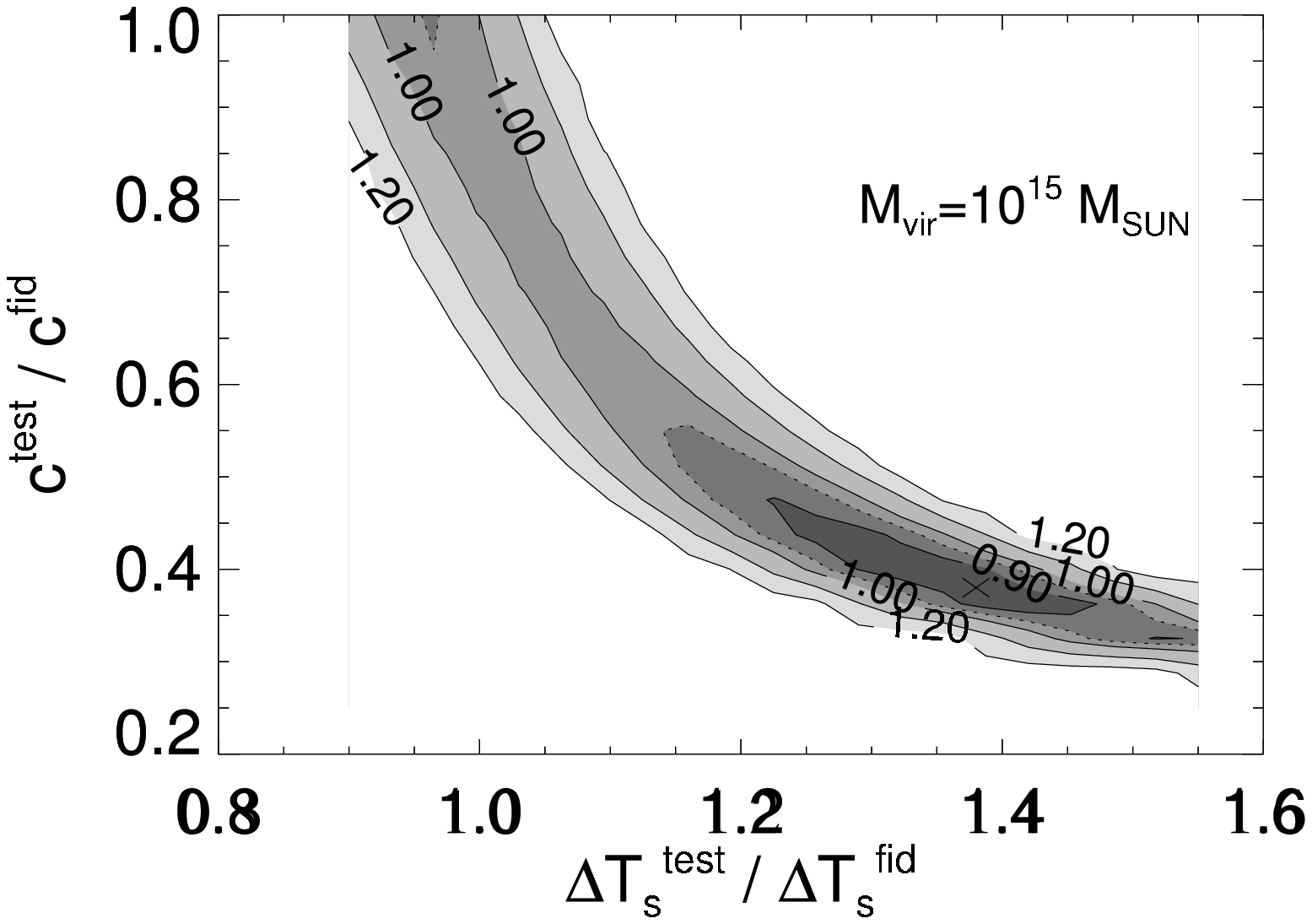}}\\
\centering\mbox{
\includegraphics[width=7cm]{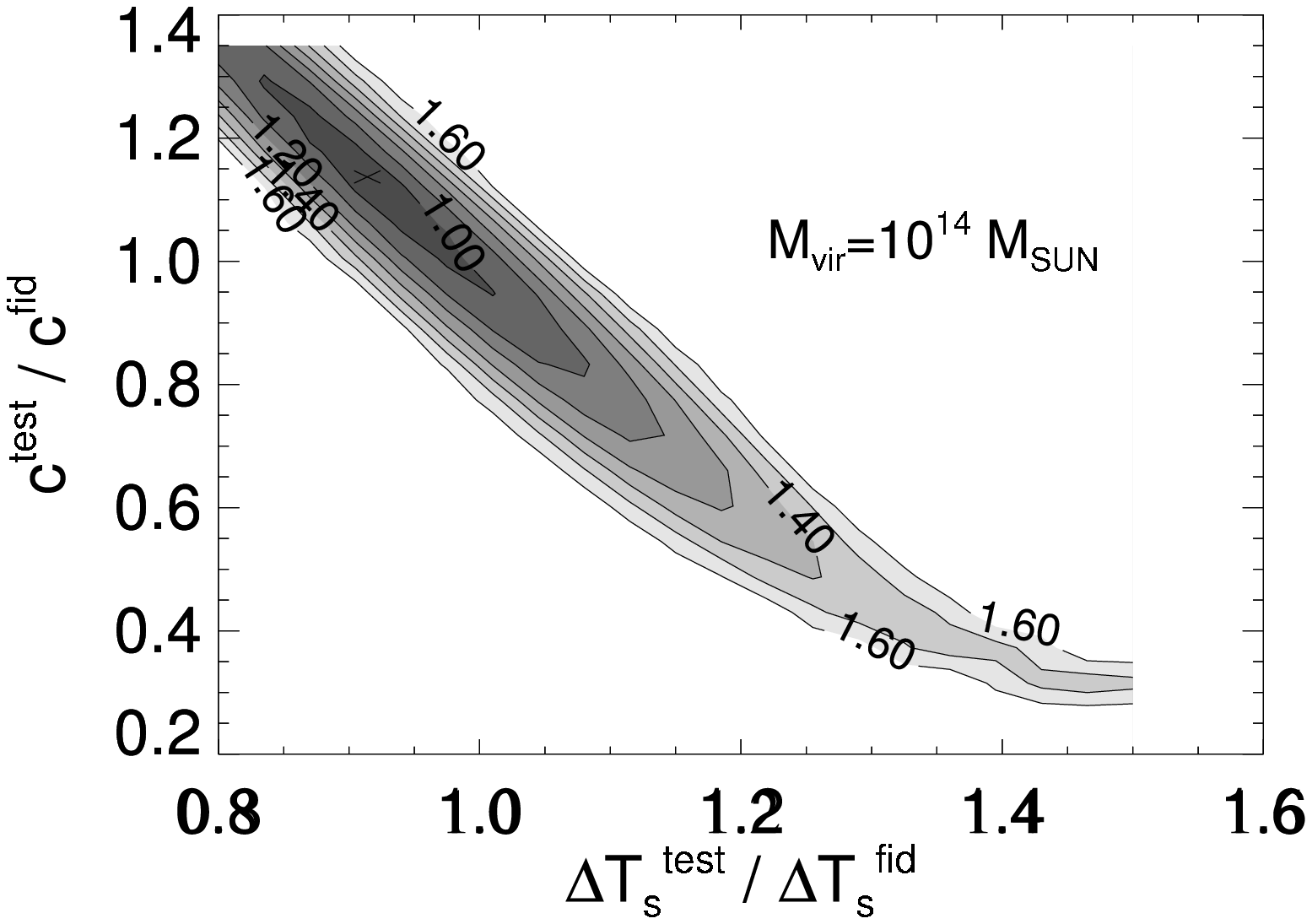}}
\caption[]{ \label{fig:H2_minimum34} The $p^{\rm test}=(\Delta
T^{\rm test}_s,c^{\rm test})$ parameter dependence of the $S/N$
ratio for detecting the virial shock of a fiducial cluster with
parameters $p^{\rm fid}=(z,\alpha,x_\mx,D)=(0.3,1.5,2c,0.5c)$ and
$M_\vir=10^{14} {\rm M_\odot}$ ({\it lower panel}) or $10^{15}
{\rm M_\odot}$ ({\it upper panel}). The fractional change in the
S/N is plotted relative to the original S/N in Model II. The
contours increase linearly in steps of 0.1 for $\delta S/N>1$, and
in steps of 0.05 for $\delta S/N<1$. The test model does not have
a virial shock, and its best--fitting parameters are biased. The
minima are marked with an X in the upper and lower panels. These
values are 0.85 for $(\Delta T^{\rm test}_0/ \Delta T^{\rm fid},
c^{\rm test}_0/c^{\rm fid})=(1.38,0.38)$ and 0.94 for $(\Delta
T^{\rm test}_0/ \Delta T^{\rm fid}, c^{\rm test}_0/c^{\rm
fid})=(0.91,1.14)$, respectively. The relatively minor decrease in
the S/N indicates that variations in $c$ and $\Delta T_s$ cannot
mimic the presence of a virial shock, i.e. that the shock remains
a distinctive feature.}
\end{figure}

The numerical results presented in the last section in
Table~\ref{tab:S/N optimal1} and Figure~\ref{fig:S/N(xmax,D,z)}
considered $p^{\rm test}$ parameter values that were equal to the
corresponding $p^{\rm fid}$ parameter values.  Here we consider
the fiducial model with $p^{\rm
fid}=(z,\alpha,x_\mx,D)=(0.3,1.5,2c,0.5c)$ and $M_\vir=10^{14}{\rm
M_\odot}$ or $10^{15}{\rm M_\odot}$. The fiducial values $(\Delta
T^{\rm fid}_s,c^{\rm fid})$ are calculated from
equations~(\ref{eq:c}) and (\ref{eq:T_s}) as before. The
parameters $p^{\rm test}$ are assumed to be the same for
$(M_\vir,z,\alpha)$ as in the fiducial model, but, in contrast
with the previous section, $(\Delta T^{\rm test}_s,c^{\rm test})$
are allowed to vary relative to the fiducial values.

The fractional deviation of the signal to noise ratio as compared
to the original value of test model I for various $p^{\rm test}$
parameters is plotted in Figure~\ref{fig:H2_minimum34}. The
$\sqrt{4/5}$ factor decrease because of the increase in the
average noise power is not included in the contour levels for
clarity. Thus, the contours have the value of 1.0 at the fiducial
parameters. The best fitting parameters $p^{\rm test}_0$
correspond to the values at the minimum of S/N.

The most important result from the analysis displayed in
Figure~\ref{fig:H2_minimum34} is that the signal to noise does not
decrease significantly, even in the best--fitting model, relative
to the values listed in the previous section (Table~\ref{tab:S/N
optimal1} and Figure~\ref{fig:S/N(xmax,D,z)}). By obtaining the
best fit for various $x_\mx$ we find that the maximum decrease is
a factor of $0.95\times\sqrt{4/5},0.9\times\sqrt{4/5}$, and
$0.5\times\sqrt{4/5}$ for $M_\vir=10^{13}{\rm M_\odot},
10^{14}{\rm M_\odot}$, and $10^{15}{\rm M_\odot}$ respectively.
The virial shock detection probabilities therefore remain
significant for $M_\vir=10^{15} {\rm M_\odot}$, but become
marginal ($S/N\lsim 10$) for low--mass clusters ($M_\vir\lsim
10^{14} {\rm M_\odot}$).

Another interesting feature shown in Figure~\ref{fig:H2_minimum34}
is that the best fitting parameters can deviate significantly from
the fiducial values. Changing $M_\vir$, $x_\mx$ and $D$ shows that
the $S/N$ of particular clusters can have saddle points and
multiple local minima in terms of the $p^{\rm test}$ parameters.
We find that this is the case, for example, for
$M_\vir=10^{14}{\rm M_\odot}$ and $1<x_\mx/c<1.5$, where the local
minimum near the fiducial values is not global, and the global
minimum $(\Delta T^{\rm test}_{s0},c^{\rm test}_0)$ instead lies
at a much larger SZ temperature and a much smaller concentration.
This behavior follows from the fact that decreasing $c$, while
fixing all other parameters, produces a steeper $\Delta T^{\rm
test}(x)$ profile on log-log scale in the outer regions, a better
approximation of the fiducial model with a virial shock. On the
other hand, the decrease in $c$ also slightly decreases the SZ
temperature in the inner regions, which can be compensated by
increasing $\Delta T_s$. Overall, a test model profile with a
smaller $T_s$ and larger $c$ comes closer to the fiducial model.

In Appendix C, we present a simple method to obtain analytical
estimates for the best--fitting $p^{\rm test}_0$ parameters of a false
test model. This approach is fully general, and gives a numerically
efficient way to compare two arbitrary models with arbitrary numbers
of parameters.  Our method is somewhat analogous to a Fisher matrix
approach.  The standard Fisher matrix treatment serves to estimate the
expected uncertainty of the parameters of a correct model.  In
contrast, the method presented in Appendix C gives an estimate of the
best--fitting parameters in an incorrect model (and does not address
the uncertainty of these best--fitting values).

\section{Parameter estimation}
\label{sec:parameters}

In the previous section, we have examined the signal to noise
ratio for distinguishing test models from the fiducial model. The
primary aim was to make a distinction between the two basic
possibilities, i.e.  whether the cluster does or does not have an
edge. A more ambitious problem is to examine the precision with
which the location and sharpness of the edge can be measured, i.e.
the precision of determining the parameters that yield the profile
$\Delta T^{\rm fid}(x)$. We calculate the uncertainty of the
parameter estimation using both a nonlinear likelihood function
directly, and with an approximate Fisher matrix method.

We adopt the fiducial model of the previous section to predict the
SZ temperature decrement profile.  As before, the fiducial model
includes an edge, and is specified by the five parameters
$M_\vir$, $z$, $\alpha$, $x_\mx$, and $D$ whose fiducial values
will here be collectively denoted by $p^{\rm fid}$.  We then
compare the temperature decrement in the fiducial model, $\Delta
T(x,p^{\rm fid})$, to two different sets of test models.  The test
models in this case are identical to the fiducial model (i.e.,
they include an edge), except we allow a subset of the parameters
to have values $p^{\rm test}\neq p^{\rm fid}$.  These two cases
are analogous to the test models I and II in the previous
sections. In the first case (Model I), we consider a set of four
unknowns $\Delta T_s$, $c$, $x_\mx$, and $D$. In the second case
(Model II), we consider only two free variables $x_\mx$ and $D$,
and fix $\Delta T_s$ and $c$ at the values calculated in the
fiducial model using equations (\ref{eq:T_s}) and (\ref{eq:c}).
The other parameters describing the cluster, $M_\vir$, $z$, and
$\alpha$, are held fixed in both cases.

Now let us suppose that a temperature decrement $\Delta T^{\rm
obs}(x)$ is measured, and with no prior information, a parameter
$p^{\rm test}$ is to be chosen that best describes the data. Due to
random noise, the measurement yields a parameter estimator with some
uncertainty. The parameter estimator and its uncertainty can be
obtained with the maximum likelihood test, following Appendix A.  The
likelihood function is
\begin{align}
L(\Delta T^{\rm obs},p)&= P(\Delta T^{\rm obs} \,|\,\Delta T(p))=
\frac{1}{(2\pi)^{K/2}(\sigma_N)^K}\times\\
&\exp\left(-\frac{K}{2A\,{\sigma_N}^2}\int\D^2x\;[\Delta T^{\rm
obs}(x)-\Delta T(x,p)]^2\right). \nonumber
\end{align}
The constant normalization pre-factor is irrelevant for the
likelihood ratio test, which we hereafter omit from the likelihood
function. The log likelihood is
\begin{equation}\label{eq:loglikelihood}
\ln L(\Delta T^{\rm obs},p)=
-\frac{K}{2A\,{\sigma_N}^2}\int\D^2x\;[\Delta T^{\rm obs}(x)-\Delta T(x,p)]^2 .
\end{equation}
where $K$ and $A$ are the total number of pixels and the surface
area of the cluster. The parameters are then chosen to maximize
$L(\Delta T^{\rm obs},p)$, or equivalently by applying a least
squares fit to the data.

The signal to noise ratio is obtained by the analysis of the
projection of the SZ observation profile on the subspace spanned
by the fiducial model profiles with arbitrary parameters. The
noise power for $N^{\rm fid}$ variable parameters is the variance
on this subspace, $N^2=N^{\rm fid}\sigma_N^2$, and the signal to
noise ratio is therefore

\begin{equation}\label{eq:false signal to noise}
\frac{S^2}{N^2}=\frac{K}{A\,N^{\rm
fid}{\sigma_N}^2}\int\D^2x\;[\Delta T^{\rm fid}(x,p^{\rm
test})-\Delta T^{\rm fid}(x,p^{\rm fid})]^2
\end{equation}

Equation (\ref{eq:false signal to noise}) measures the extent that
the parameter estimator is $p^{\rm test}$ instead of $p^{\rm
fid}$, the true (and maximum likelihood) value.
$S^2/\sigma_N^2$ follows $\chi^2$-statistics with $N^{\rm fid}$
degrees of freedom, leading to $S^2/N^2=1\pm\sqrt{2/N^{\rm fid}}$
for $1\sigma$ errors.
Given a true parameter set $p^{\rm fid}$, the region within $2\sigma$
confidence (95\%) for example is the set of $p^{\rm test}$ values for
which $S^2/N^2<1+2\sqrt{2/N^{\rm fid}}$.  The uncertainty of the
parameter estimation can therefore be read off directly from
equation~(\ref{eq:false signal to noise}).

\begin{figure}
\centering\mbox{
\includegraphics[width=7cm]{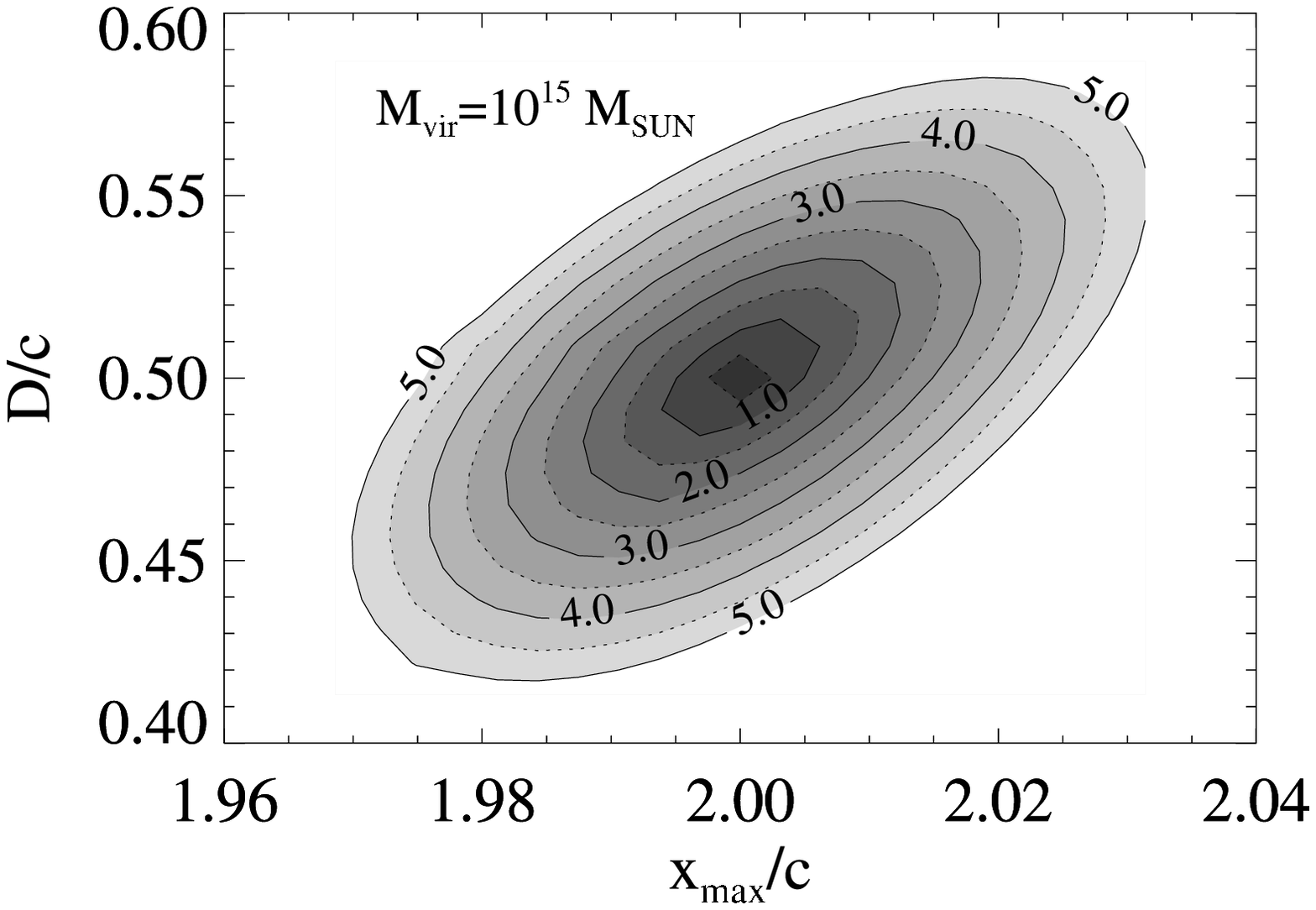}}\\
\centering\mbox{
\includegraphics[width=7cm]{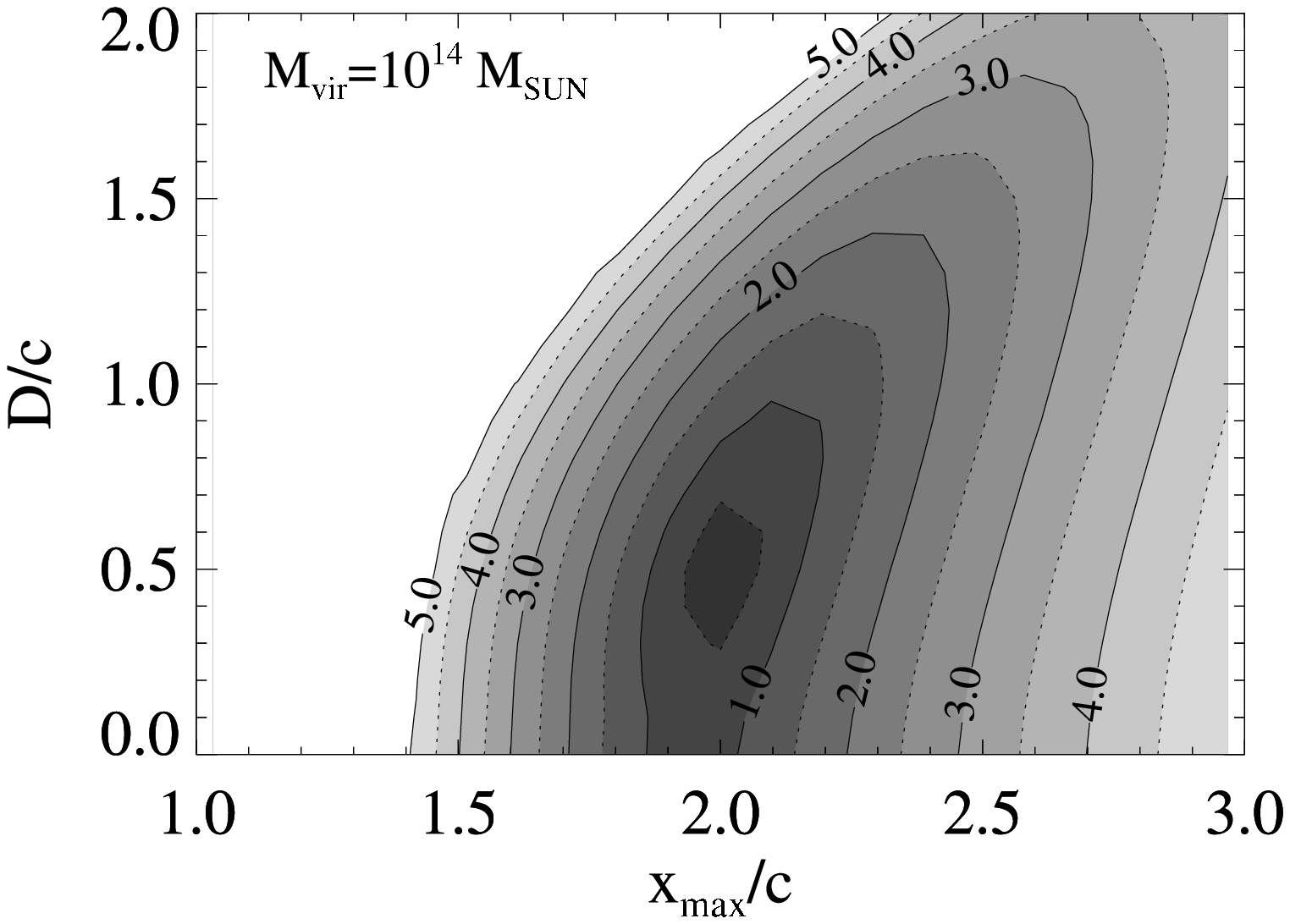}}
\caption[]{ \label{fig:parameter variance12}The $(x^{\rm
test}_{\mx}, D^{\rm test})$ parameter dependence $\chi^2$-contours
for Model II with $(z,\alpha,x_\mx,D)=(0.3,1.5,2c,0.5c)$, using
$M_\vir=10^{14}{\rm M_\odot}$ ({\it lower panel}) and
$M_\vir=10^{15}{\rm M_\odot}$ ({\it upper panel}). The contours
increase linearly in steps of $0.5\sigma_N$.}
\end{figure}

Figure~\ref{fig:parameter variance12} shows contour diagrams at
fixed values of this uncertainty in Model II, assuming the
fiducial parameters $p^{\rm
fid}=(z,\alpha,x_\mx,D)=(0.3,1.5,2c,0.5c)$ with $M=10^{14}{\rm
M_\odot}$ (upper panel) or $M=10^{15}{\rm M_\odot}$ (lower panel).
The contour plots allow $x_\mx$ and $D$ to vary, while all other
parameters are held fixed at the fiducial values.

Figure~\ref{fig:parameter variance12} depicts the likelihood
contours for Model II with the direct evaluation of the
$S/\sigma_N$. An estimate for these contours is provided by the
Fisher information matrix,

\begin{align}
F_{i,j}(p)&=-\left\langle\frac{\partial^2 \ln
L(\Delta T^{\rm obs},p)}{\partial
p_i\partial p_j}\right\rangle_{\Delta T^{\rm obs}} \\
&= \frac{K}{A\,\sigma_N^2}\int \D^2x\; \frac{\partial \Delta T(x,p^{\rm fid})}{\partial p_i}
\frac{\partial \Delta T(x,p^{\rm fid})}{\partial p_j}.
\end{align}

Assuming that the likelihood distribution is Gaussian near the
peak likelihood, we can use confidence limits for Gaussian
statistics (i.e., $\chi^2$) to obtain $68\%$ and $95\%$ confidence
regions.  The Fisher matrix greatly speeds up the computations,
and allows us to consider additional free parameters (Model I).

The minimum expected variance is related to the Fisher matrix by
the Cramer-Rao bound \citep{teg97}
\begin{equation}\label{eq:Cramer-Rao}
(\Delta p_j)^2 \geq (F^{-1})_{jj}
\end{equation}
where equality holds if the distribution is well approximated by
a Gaussian distribution.

Tables~\ref{tab:parameter variance2} and \ref{tab:parameter
variance1} show the 68\% and 95\% significance uncertainties for
Models I and II, as calculated from the Fisher matrix.  Generally,
increasing the number of free variables can increase the
uncertainty of the original variables, since a model with
additional free parameters can better mimic the fiducial model for
a given mock observation.  However, our calculations indicate that
this is not the case in our case: the uncertainties on $x_\mx$ and
$D$ are nearly the same in Models I and II.  This is perhaps
unsurprising, since one would not expect that a change in $c$ or
$\Delta T_s$ (the additional free parameters in Model I) can mimic
changes in either $x_\mx$ and $D$.  Indeed, the Fisher matrix is
decoupled in the corresponding subspaces, as the $(x_\mx,\Delta
T_s)$, $(x_\mx,c)$, $(D,\Delta T_s)$, $(D,c)$ components are
negligible compared to the diagonal elements. The uncertainties in
the parameter estimators are $(\delta c/c,\delta \Delta T_s/\Delta
T_s,\delta x_\mx/x_\mx,\delta D/x_\mx)=(4\%,6\%,9.5\%,25\%)$ for
$(M_\vir,z,\alpha,c,\Delta
T_s,x_\mx,D)=(10^{14}M_\odot,0.3,1.5,2.9,1.2\mK,2c,0.5c)$.
Therefore $c$, $\Delta T_s$, and $x_\mx$ will be precisely
obtained with ALMA for the majority of the clusters, while similar
precision for $D$ is possible for only massive clusters. We note
that in principle, it is possible to deduce the value of $\alpha$
directly from the $c(\alpha)$ relation.

Figure~\ref{tab:parameter variance2} provides confidence in our
conclusions from the Fisher matrix method: the $1\sigma$ contours
in $(x_\mx,D)$ are well-approximated by ellipses, if the fiducial
values obey $D-\Delta D>0$, e.g. $D>0.1$, $x_\mx<1.7c$, and
$M_\vir\geq 10^{14}{\rm M_\odot}$. Therefore, in these cases, the
inequality (\ref{eq:Cramer-Rao}) assumes equality for the
parameter uncertainty. We find that the parameter distribution
around $D\approx 0$ is distorted, and higher confidence level
contours for arbitrary $D$ are banana shaped in the $x_\mx-D$
plane.

\begin{deluxetable}{cccccccc}
\tablecolumns{8} \tablewidth{0pt}
\tablecaption{\label{tab:parameter variance2} Parameter
uncertainty ($z=0.3$, Model II)} \tablehead{ \colhead{$M_\vir$} &
\colhead{$\alpha$} & \colhead{$\Theta$} & \colhead{c} &
\colhead{$\Delta T_s$} & \colhead{$x_\mx$} &
\colhead{$D$} & \colhead{S} \\
\colhead{$[M_\odot]$}   &  \colhead{} & \colhead{[arcmin]} &
\colhead{}  & \colhead{[-muK]} & \colhead{$[c]$} & \colhead{$[c]$}
& \colhead{\%}} \startdata
      $10^{13}$ &  1   & 1.8   &  7.9 & 23 &   2  &   0.5 & -\\
      - & - & -      &  - & -  &    6.5  & 17 &  68\\
      - & - & -      &  - & -  &     13  &  34 & 95 \\

      $10^{13}$ &  1.5 & 1.7 &  4.6 & 170 &   2 &   0.5& - \\
      - & - & -       &  - & -  &   6.1    &  16  & 68\\
      - & - & -       &  - & -  &   12    &  32  & 95\\
      $10^{14}$ &  1   & 3.8   &  5.0 & 140   &   2  &   0.5& - \\
      - & - & -        &  - & -   &  0.20 &   0.55 & 68\\
      - & - & -        &  - & -   &  0.41  &   1.1 & 95\\
      $10^{14}$ &  1.5 & 3.8   &  2.9 &1200 &   2  &   0.5 & -\\
      - & - & -      & - & - &  0.19  &   0.52 &68\\
      - & - & -      & - & - &  0.38  &   1.04 &95\\
      $10^{15}$ &  1   & 8.1   &  3.1 & 860 &   2  &   0.5 & -\\
      - & - & -     &  - & - &  0.006  &   0.018 & 68\\
      - & - & -     &  - & - &  0.013  &   0.036 & 95\\
      $10^{15}$ &  1.5  & 8.1   &  1.8 &7900&   2  &   0.5 &-\\
      - & - & -       & -  & - &  0.006 &  0.017 & 68\\
      - & - & -       & -  & - &  0.012 &   0.035 & 95 \\
\enddata
\end{deluxetable}

\begin{deluxetable}{cccccccc}
\tablecolumns{8} \tablewidth{0pt}
\tablecaption{\label{tab:parameter variance1} Parameter
uncertainty ($z=0.3$, Model I)} \tablehead{ \colhead{$M_\vir$} &
\colhead{$\alpha$} & \colhead{$\Theta$} & \colhead{c} &
\colhead{$\Delta T_s$} & \colhead{$x_\mx$} &
\colhead{$D$} & \colhead{S} \\
\colhead{$[M_\odot]$}   &  \colhead{} & \colhead{[arcmin]} &
\colhead{}  & \colhead{[-muK]} & \colhead{$[c]$} & \colhead{$[c]$}
& \colhead{\%}} \startdata
      $10^{13}$ &   1   & 1.8   &  7.9 & 23  &   2  &   0.5 & -\\
      - & - & -      &  1.0 & 15  &    6.5  & 17 &  68\\
      - & - & -      &  2.1 & 29  &    13  &  34  & 95 \\
      $10^{13}$ &   1.5 & 1.7   &  4.6 & 170  &   2 &   0.5& - \\
      - & - & -       &  0.83 & 132  &     6.1  &   16 & 68\\
      - & - & -      &  1.7 & 263  &     12 &   32 & 95\\
      $10^{14}$ &  1   & 3.8   &  5.0 & 140    &   2  &   0.5& - \\
      - & - & -        &  0.15 & 6.4  &    0.21  &   0.55 & 68\\
      - & - & -        &  0.30 & 13   &    0.41  &   1.09 & 95\\
      $10^{14}$ &   1.5 & 3.8   &  2.9 &1200   &   2  &   0.5 & -\\
      - & - & -      & 0.11 & 67  &   0.19  &   0.52 &68\\
      - & - & -      &  0.22 & 135 &   0.39  &   1.05 &95\\
      $10^{15}$ &  1   & 8.1   &  3.1 & 860 &   2  &   0.5 & -\\
      - & - & -     &  0.013 & 2.2 &   0.007  &   0.018 & 68\\
      - & - & -     &  0.026 & 4.4 &   0.013  &   0.036 & 95\\
      $10^{15}$ &  1.5  & 8.1   &  1.8 &7900 &   2  &   0.5 &-\\
      - & - & -       & 0.011  & 27   &  0.006 &  0.017 & 68\\
      - & - & -       & 0.023   & 55  &   0.013  &   0.035 & 95 \\
\enddata
\end{deluxetable}

Our calculations show that increasing $x_\mx$ increases the
uncertainties $\Delta x_\mx$ and $\Delta D$, while increasing $D$ only
slightly increases $\Delta x_\mx$ and leaves $\Delta D$ unchanged.  It
is clear that the choice of $\alpha$ does not alter the $\Delta x_\mx$
and $\Delta D$ uncertainties. Changing $z$ yields a maximum in $\Delta
x_\mx$ and $\Delta D$ around $z=0.4$. We find that the parameter
uncertainties decrease by a factor of $\sim 10$ at $z\approx 0$ or
$z\approx 3$, compared to the $z=0.4$ value.


\section{Conclusions}
\label{sec:conclusions}

We have shown that the forthcoming Atacama Large Millimeter Array
(ALMA) is well--suited for studies of the intra--cluster medium
density distribution using the Sunyaev-Zel'dovich (SZ) effect. The
angular beam diameter and sensitivities are predicted to reach $2"$
and $10\muK$ for this system, which exceed present detector
resolutions by more than two orders of magnitude. The SZ decrement
profile of a rich galaxy cluster, observed at 100 GHz for $\sim$30
hours, will have a sufficiently high signal-to-noise ratio to make
inferences about the gas distributions (and, by inference, about
the dark matter distributions).

We solved the equations of hydrostatic equilibrium for the
self--similar gas density in the dark matter background. The dark
matter profile was parameterized by its inner slope $\alpha$, and
every calculation was evaluated for the two values common in the
literature, i.e. $\alpha=1$ and 1.5.  Within this framework, the
density as a function of radial distance is small but nonzero,
even in the outer regions. As a modification, we introduced a
linear cutoff in the density profile, where the density falls to
zero within a finite radial distance. This assumption mimics the
presence of strong virial shocks, predicted in cosmological
structure formation theories (see \citealt{bert85} and
\citealt{Bagchi}).

We have calculated the SZ decrement profiles of clusters with and
without the presence of a sharp density cutoff near the virial
radius. We assumed $10\muK$ flat-power Gaussian noise.  We set up
the likelihood function and defined the decision rule to see
whether our fiducial model with a virial shock can be
distinguished from a test model without a virial shock. It is
important to emphasize that this test is practically
model--independent: it is only weakly sensitive to the central
regions, where cluster models tend to differ, with most of the
sensitivity arising from the outer regions, where the various
cluster models in the literature are very similar. We calculated
the signal to noise ratio in two ways, with either four (Model I)
or two (Model II) free parameters. These parameters are the radial
distance to the virial shock,
$x_\mx$, the radial thickness of the cutoff, $D$, the
concentration parameter, $c$, and the temperature decrement scale,
$\Delta T_s$, for Model I. For Model II, $c$ and $\Delta T_s$ were
calculated from theory assuming no scatter around the correct
values. Other parameters, such as the virial mass, $M_\vir$, the
redshift, $z$, and the logarithmic slope of the inner dark matter
profile, $\alpha$, can, in principle, be obtained from independent
measurements, and were not varied. The signal to noise ratios
agree to within 20\% in Model I and Model II. The overall value is
S/N=470 for a rich cluster of virial mass $10^{15} M_\odot$ and we
find S/N=16 for a lower--mass cluster with $10^{14} M_\odot$.

We also considered a more realistic case, in which the parameters
of the cluster with and without a virial shock were fitted
independently for a given measurement. We find a large systematic
bias in the parameters $c$ and $\Delta T_s$ for the model without
a virial shock, provided that the cluster does have a virial shock
within approximately two virial radius. Nevertheless, the signal
to noise ratio decreases only by $50\%$ at most for detecting the
virial shock. Therefore, the virial shocks remain detectable for
$M\gsim 10^{14}{\rm M_\odot}$ clusters.

Finally, we examined the precision to which the parameters of the
virial shock could be obtained. We find the typical values for the
precisions for low--mass ($10^{14}~{\rm M_\odot}$) and rich
($10^{15}~{\rm M_\odot}$) clusters to be $ (\delta c/c, \delta
\Delta T(0)/\Delta T(0), \delta x_\mx/c,\delta
D/x_\mx)=(4\%,6\%,9.5\%,25\%)$ and $(0.6\%,0.3\%,0.3\%,0.9\%)$,
respectively.  These results assume the cluster is at $z=0.3$, and
it is surrounded by a virial shock at twice the virial radius
($x_\mx=2c$) that has a finite width $D=0.5c$. If the shock was
located closer in ($x_\mx=c$) and was sharper ($D=0.01c$), these
uncertainties would improve to $(3.7\%,6.1\%,2.8\%,12\%)$ and
$(0.6\%,0.4\%,0.1\%,0.4\%)$, respectively.  Since there is a
one--to--one theoretical correspondence between $c$ and $\alpha$,
with $c(\alpha=1)/c(\alpha=1.5)=1.7$, SZ measurements should help
in determining the value of $\alpha$.  This should serve as a
valuable diagnostic of the inner gas profile; in particular, of
the presence of any excess (non--gravitational) entropy that could
strongly modify the inner profile.

Our analysis is based on a simplified, spherically symmetric model
for the cluster.  In these models, the virial shock appears in the
SZ surface brightness maps as a full (2$\pi$ azimuthal angle)
ring--like feature.  In reality, the gas--infall into the cluster
potential should be inhomogeneous and anisotropic. As a result,
the radial location and the strength of the virial shock will vary
in different directions away from the cluster's center
\citep{evrard90,bn98}.  We partially accounted for this effect by
allowing the virial shock to have a finite width ($D$), which
``smears out'' the ring--like feature -- mimicking the effect of
projecting shocks of varying radial locations and strengths.
Simulations also show that gas is falling into the cluster
potential along overdense filaments, and that the strongest shocks
occur along these filaments.  If the projected ring--like feature
extended over only a fraction $f$ of the full $2\pi$ azimuthal
angle (corresponding to the projected areas of the filaments),
then our predicted $S/N$ for detecting the virial shock would be
further reduced by a factor $1/\sqrt{f}$.  The virial shock will
nevertheless remain detectable for a $M\sim 10^{15}{\rm M_\odot}$
cluster at a significance of $S/N\gsim 10$, unless
$f\lsim10^{-3}$.

Our results suggest that SZ decrement measurements with ALMA can
reveal virial shocks around galaxy clusters, and can determine the
location and size of these shocks with high precision.  This will
provide a unique constraint on theories of large--scale structure
formation.

\acknowledgments We thank Gil Holder, Greg Bryan, David Helfand,
and S\'andor Moln\'ar for useful discussions. We acknowledge
support from OTKA through grant nos. T037548, T047042, and
T047244.


\section*{APPENDIX A: DISTINGUISHING MODELS WITH THE OPTIMAL FILTER}

In this appendix, we give a formal summary of the treatment of the
following general problem.  Consider a system that obeys laws that
predict a set of observables ${\bf h}^{\rm fid}$. What is the
significance at which a false hypothesis, predicting a different set
of observables, ${\bf h}^{\rm test}$, can be rejected, given the
presence of Gaussian random noise?

A measurement of the observable in $x_K$ distinct directions yields a
discrete sample $y_k$ where $k\in[1,K]$.  The set $\{y_k\}$ is an
element of a $K$-dimensional vector space and will be denoted by
${\bf y}$. Similarly, let ${\bf h}^{\rm fid}$ and ${\bf h}^{\rm test}$
denote the discrete sample of the hypotheses functions $h^{\rm
fid}(x)$ and $h^{\rm test}(x)$. In our example, ${\bf y}$ denotes a
measured SZE temperature profile, whereas ${\bf h}^{\rm fid}$ and
${\bf h}^{\rm test}$ denote the discrete temperature profiles predicted
in models with and without a cutoff. The vectors ${\bf h}^{\rm fid}$
and ${\bf h}^{\rm test}$ depend on the parameters describing the
cluster and the cutoff. We now derive the signficance (or ``signal to
noise ratio'') of distinguishing between two models.

If the real signal arriving to the detector was $s(x)$, then the
detector measures the data

\begin{equation}
{\bf y}={\bf s}+{\bf n}
\end{equation}
where ${\bf n}$ is a random variable, corresponding to the noise.
Let us assume white Gaussian noise of variance $\sigma_N^2$.
In this case, the probability of detecting ${\bf y}$, given that
the incoming signal is ${\bf s}$, is
\begin{equation}\label{eq:P(y|s)}
P({\bf y}\,|\, {\bf s})= \frac{1}{(2\pi)^{K/2}(\sigma_N)^K}
\exp\left[-\frac{({\bf y}-{\bf s})^2}{2\sigma_N^2} \right] ,
\end{equation}
where the arithmetic (difference and scalar product) of the
K-dimensional vectors was used, i.e.
\begin{equation}
({\bf y}-{\bf s})^2=\sum_{k=1}^K (y_k - s_k)^2 .
\end{equation}

Two hypotheses can be distinguished using $P({\bf s}\,|\,{\bf y})$,
which can be obtained from $P({\bf y}\,|\, {\bf s})$ and the a--priori
probabilities $P({\bf s})$ by applying the Bayes theorem for
conditional probabilities. Most common decision rules, such as the
maximum posteriori, the Neyman-Pearson, or the minimax decisions
(e.g. \citealt{whalen}) involve constraints on the likelihood ratio,
\begin{equation}\label{eq:L(y)def}
L({\bf y})= \frac{P({\bf y}\,|\,{\bf h}^{\rm
test})}{P({\bf y}\,|\,{\bf h}^{\rm fid})} .
\end{equation}

Substituting (\ref{eq:P(y|s)}) in (\ref{eq:L(y)def}) the
log--likelihood becomes
\begin{equation}\label{eq:lnL(y)}
\ln L({\bf y})=\frac{1}{\sigma_N^2}({\bf h}^{\rm test}-{\bf h}^{\rm
fid})\cdot({\bf y}-{\bf h}^{\rm fid}) -
\frac{1}{2\sigma_N^2}({\bf h}^{\rm test}-{\bf h}^{\rm fid})^2
\end{equation}
The decision is made in favor of the test hypothesis
if the likelihood ratio exceeds a given threshold.
The likelihood depends on the ${\bf y}-{\bf h}^{\rm fid}$
component along ${\bf h}^{\rm test}-{\bf h}^{\rm fid}$. The term
${\bf h}^{\rm test}-{\bf h}^{\rm fid}$ is referred to as the
matched filter, which is the optimal filter for the distinction of
the two hypotheses in white Gaussian noise.

Since the noise distribution $P({\bf n})$ is spherically symmetric
in the K-dimensional vector space, it has the same power
$\sigma_N^2$ along any basis. This implies that the noise power
for the likelihood detection rule is the noise power for a single
bin, i.e. $N^2=\sigma_N^2$. Thus increasing the sample size
increases the signal power, but leaves the relevant noise
contribution the same.

The signal power is
\begin{equation}\label{eq:signal1}
S^2=({\bf h}^{\rm test}-{\bf h}^{\rm fid})^2=\sum_k(h^{\rm
test}_{k}-h^{\rm fid}_{k})^2=\frac{1}{\Delta x^2}\int \D^2 x\;
\Delta h(x)^2 ,
\end{equation}
where $\Delta h(x)=h^{\rm test}(x)-h^{\rm fid}(x)$ and
$\Delta x^2$ denotes the area enclosed by the neighboring $x_k$
points. We have assumed that the resolution of the measurement is
fine enough to approximate the sum with the integral. Therefore,
the signal to noise ratio is
\begin{equation}\label{eq:signal to noise simple}
\frac{S^2}{N^2}=\frac{({\bf h}^{\rm test}-{\bf h}^{\rm
fid})^2}{\sigma_N^2}.
\end{equation}
The signal to noise ratio is an important measure of the
significance of the test. Let us define $\rho$ as a
one-dimensional random variable by
\begin{equation}
\rho(\bf y)=\frac{({\bf h}^{\rm test}-{\bf h}^{\rm fid})\cdot({\bf
y}-{\bf h}^{\rm fid})}{||{\bf h}^{\rm test}-{\bf h}^{\rm
fid}||\,\sigma_N}
\end{equation}
for which $P(\rho\,|\, {\bf s}={\bf h}^{\rm fid})$ is a standard
normal distribution, whereas $P(\rho\,|\, {\bf s}={\bf h}^{\rm
test})$ is a normal distribution with mean $S/N$ and unit
variance. Thus, if the fiducial model is true, $\rho^2$ follows
$\chi^2$ statistics, and the significance of the test can be
obtained by comparing $S^2/N^2$ with the $\chi^2$ confidence
levels.

In our case, for a detector with angular resolution $\Delta\phi$ and a
cluster of apparent angular virial radius $\Theta$ the surface element
is $\Delta x^2=A/K$, where $K=(x^{\rm
test}_\mx\Theta)^2/(c\Delta\phi)^2$ is the total number of pixels and
$A=\pi {x^{\rm test}_\mx}^2$ is the surface area of the
cluster. Therefore,
\begin{equation}\label{eq:S/N optimal0}
\frac{S}{N} = \frac{\Theta}{\sqrt{\pi}\,c\,\Delta \phi\, \sigma_N}
\sqrt{\int \D^2x\, [h^{\rm test}(x)-h^{\rm fid}(x)]^2}.
\end{equation}
is the signal to noise ratio for the decision between the two
hypotheses.


\section*{APPENDIX B: DISTINGUISHING MODELS WITH FREE PARAMETERS}

Here we address the following problem, which is a generalized
version of the problem posed in Appendix A above.  Consider again
a system that obeys laws that predict a set of observables ${\bf
h}^{\rm fid}$. What is the significance at which a false
hypothesis, predicting a different set of observables, ${\bf
h}^{\rm test}$, can be rejected, given the presence of a Gaussian
random noise, {\it if the parameters of the false test model are
not known a priori and can be freely adjusted}?

Assume that the original signal (i.e. without noise) is ${\bf
h}^{\rm fid}$ with a set of $N^{\rm fid}$ parameters, $p^{\rm
fid}_0$,
and the false hypotheses is described by $N^{\rm test}$
parameters, $p^{\rm test}$. The measurement
\begin{equation}\label{eq:y_distribution}
{\bf y}={\bf h}^{\rm fid}(p^{\rm fid}_0)+{\bf n}
\end{equation}
can be used to give an estimate of $p^{\rm fid}_0$. As in Appendix
A, ${\bf y}$ is the collection of $K$ observables (e.g. the SZ
brightness of the $K$ pixels), and ${\bf n}$ is a $K$-dimensional
Gaussian random variable. Denote the estimated parameter of ${\bf
h}^{\rm fid}$ by $p^{\rm fid}$ and the estimated parameter of
${\bf h}^{\rm test}$ by $p^{\rm test}$. Once the parameters have
been obtained, ${\bf h}^{\rm fid}(p^{\rm fid})$ and ${\bf h}^{\rm
test}(p^{\rm test})$ can be fixed at the corresponding values.
Then the likelihood ratio for fixed parameters can be used,
according to equation~(\ref{eq:lnL(y)}), and the decision is made
in favor of the test model rather than the fiducial model exactly
if the likelihood is above a given threshold.

Since $p^{\rm fid}$ is obtained by minimizing $({\bf y}-{\bf
h}^{\rm fid}(p^{\rm fid}))^2$, the distribution of ${\bf h}^{\rm
fid}(p^{\rm fid})$ from (\ref{eq:y_distribution}) is
\begin{equation}\label{eq:h1(p1)}
{\bf h}^{\rm fid}(p^{\rm fid}) = {\bf h}^{\rm fid}(p^{\rm fid}_0)
+ \bm{\Delta}_1 ,
\end{equation}
where $p^{\rm fid}_0$ is the real parameter of the signal without
the noise and $\bm{\Delta}_1$ is an $N^{\rm fid}$ dimensional
Gaussian random variable. \footnote{Note that $\bm{\Delta}_1$ is a
Gaussian random variable in terms of the set of parameters $p^{\rm
fid}$ only if the various choices for these parameters span a flat
(linear) subspace. This is an adequate approximation, provided
that the radius of curvature is much less than the r.m.s. variance
of the parameters, $(\sqrt{N^{\rm fid}}\sigma_N)$.} $\bm{
\Delta}_1$ has zero mean and variance $N^{\rm fid}\sigma_N^2$.
Similarly, $p^{\rm test}$ is obtained by minimizing $({\bf y}-{\bf
h}^{\rm test}(p^{\rm test}))^2$. The distribution of ${\bf h}^{\rm
test}(p^{\rm test})$ is
\begin{equation}\label{eq:h2(p2)}
{\bf h}^{\rm test}(p^{\rm test}) = {\bf h}^{\rm test}(p^{\rm
test}_0) + \bm{\Delta}_2
\end{equation}
where $p^{\rm test}_0$ is the $p^{\rm test}$ parameter value at
which $({\bf h}^{\rm fid}(p^{\rm fid}_0)-{\bf h}^{\rm test}(p^{\rm
test}))^2$ is minimal. $\bm{\Delta}_2$ is defined by
equation~(\ref{eq:h2(p2)}), which is an $N^{\rm test}$-dimensional
Gaussian random variable, with zero mean and variance $N^{\rm
test}\sigma_N^2$. Note that $\bm{\Delta}_1$ and $\bm{\Delta}_2$
are strongly correlated since both values are derived from the
same measurement ${\bf y}$.

The signal power
\begin{equation}
S^2=({\bf h}^{\rm test}-{\bf h}^{\rm fid})^2
\end{equation}
is therefore a random variable. Let us define the empirical, most
probable\footnote{The estimated parameters have a Gaussian
distribution around $p^{\rm fid}_0$ and $p^{\rm test}_0$. The term
''most probable'' refers to the parameter distribution not the
signal power distribution.} and expected signal powers by
\begin{align}
S_{\rm emp}^2&=({\bf h}^{\rm test}(p^{\rm test})-{\bf h}^{\rm fid}(p^{\rm fid}))^2 \\
\label{eq:S_0} S_0^2&=({\bf h}^{\rm test}(p^{\rm test}_0)-{\bf
h}^{\rm fid}(p^{\rm fid}_0))^2 = \min_{p^{\rm test}} ({\bf h}^{\rm
test}(p^{\rm test})-{\bf h}^{\rm fid}(p^{\rm fid}_0))^2
\\
S_{\rm exp}^2&=\langle S_{\rm emp}^2 \rangle _{p^{\rm test},p^{\rm
fid}} .
\end{align}

The expected signal power can be written in terms of $S_0$ using
eqs. (\ref{eq:h1(p1)}) and (\ref{eq:h2(p2)})
\begin{align}
S_{\rm exp}^2 &=\langle[{\bf h}^{\rm test}(p^{\rm test}_0)-{\bf
h}^{\rm fid}(p^{\rm fid}_0) + \bm{\Delta}_2 -
\bm{\Delta}_1 ]^2\rangle \\
&= S_0^2+ \langle(\bm{\Delta}_2 -
\bm{\Delta}_1 )^2 \rangle\\
&=S_0^2+ \langle\bm{\Delta}_2^2\rangle +
\langle\bm{\Delta}_1^2\rangle - 2 \langle \bm{\Delta}_1
\bm{\Delta}_2\rangle\\
&=S_0^2 +  (N^{\rm fid} + N^{\rm test})\sigma_N^2 - 2 \langle
\bm{\Delta}_1 \bm{\Delta}_2\rangle. \label{eq:signal exact}
\end{align}
In particular if $N^{\rm fid}=N^{\rm test}=1$, then $\langle
\bm{\Delta}_1 \bm{\Delta}_2\rangle = \cos(\phi)\sigma_N^2/2 $.
This term can be obtained, for given ${\bf h}^{\rm fid}$ and ${\bf
h}^{\rm test}$ hypothesis functions by
\begin{equation}
\cos(\phi) = \frac{\left|\frac{\D {\bf h}^{\rm fid}(p^{\rm
fid}_0)}{\D p^{\rm fid}_0}\cdot \frac{\D {\bf h}^{\rm test}(p^{\rm
test}_0)}{\D p^{\rm test}_0}\right|}{\left|\frac{\D {\bf h}^{\rm
fid}(p^{\rm fid}_0)}{\D p^{\rm fid}_0}\right| \left|\frac{\D
{\bf h}^{\rm test}(p^{\rm test}_0)}{\D p^{\rm test}_0}\right|} .
\end{equation}
For arbitrary numbers of parameters, the expected value of
$\langle \bm{\Delta}_1 \bm{\Delta}_2\rangle$ has a more
complicated algebraic form, but can always be obtained for a
specific choice of ${\bf h}^{\rm fid}$ and ${\bf h}^{\rm test}$
functions. For simplicity we shall restrict only to the inequality
\begin{equation}
\langle \bm{\Delta}_1 \bm{\Delta}_2\rangle \geq 0 ,
\end{equation}
leading to
\begin{equation}\label{eq:signal}
S_0^2 \leq S_{\rm exp}^2 \leq S_0^2 + (N^{\rm fid} + N^{\rm
test})\sigma_N^2 .
\end{equation}

The noise power for choosing between ${\bf h}^{\rm fid}$ and
${\bf h}^{\rm test}$ (with an arbitrary $p^{\rm fid}$ or $p^{\rm
test}$) is the variance of the measurement along ${\bf h}^{\rm
test}(p^{\rm test})-{\bf h}^{\rm fid}(p^{\rm fid})$, since the
likelihood ratio (\ref{eq:lnL(y)}) depends on only this component.
Therefore
\begin{align}
N_{\rm exp}^2&=\left< \left[\frac{{\mathbf h}^{\rm test}(p^{\rm
test})-{\bf h}^{\rm fid}(p^{\rm fid})}{||{\bf h}^{\rm test}(p^{\rm
test})-{\bf h}^{\rm fid}(p^{\rm fid})||}
\cdot ({\bf y}-{\bf h}^{\rm fid}(p^{\rm fid})) \right]^2\right>\\
&= \sigma_N^2 + \left< \left[\frac{{\bf h}^{\rm test}(p^{\rm
test})-{\bf h}^{\rm fid}(p^{\rm fid})}{||{\bf h}^{\rm test}(p^{\rm
test})-{\bf h}^{\rm fid}(p^{\rm fid})||} \cdot \bm{\Delta}_1
\right]^2\right> .
\end{align}
For large signal to noise ratios this can be approximated to
lowest order in $\bm{\Delta}_1$.
\begin{align}
N_{\rm exp}^2&= \sigma_N^2 + \left< \left[\frac{{\bf h}^{\rm
test}(p^{\rm test}_0)-{\bf h}^{\rm fid}(p^{\rm fid}_0)} {||{\bf
h}^{\rm test}(p^{\rm test}_0)-{\bf h}^{\rm fid}(p^{\rm fid}_0)||}
\cdot \bm{\Delta}_1
\right]^2\right>\\
&=\sigma_N^2 + \langle(\cos\theta)^2\rangle \sigma_N^2 ,
\end{align}
where $\theta$ is the angle between the vector ${\bf h}^{\rm
test}(p^{\rm test}_0)-{\bf h}^{\rm fid}(p^{\rm fid}_0)$ and ${\bf
h}^{\rm fid}(p^{\rm fid})-{\bf h}^{\rm fid}(p^{\rm fid}_0)$. In
general for any $({\bf h}^{\rm fid},{\bf h}^{\rm test})$
\begin{equation}\label{eq:noise}
\sigma_N^2 \leq N_{\rm exp}^2 \leq 2\sigma_N^2 .
\end{equation}
Equation (\ref{eq:noise}) is a natural consequence of the fact
that there are two sources of uncertainties for the hypothesis
test, corresponding to the uncertain estimates of the parameters
of the fiducial and test models. In principle, these uncertainties
can add up constructively to create a $2\sigma_N^2$ variance in
the likelihood ratio, but the noise power is at least the noise of
the fixed models of Appendix A, $1\sigma_N^2$.

In particular, for the SZE brightness hypotheses, the test model
is a special case of the fiducial model, with the fiducial model
having two additional parameters describing the cutoff at the
virial shock. In this case for $N^{\rm fid}=4$ and $N^{\rm
test}=2$, it can be shown that
$\langle(\cos\theta)^2\rangle\approx 1/4$ independent of $p^{\rm
fid}_0$.
Therefore
\begin{equation}\label{eq:noise exact}
N_{\rm exp}^2=\frac{5}{4}\sigma_N^2 .
\end{equation}

Although the expected signal to noise ratio can be calculated
explicitly for a given parameter choice using
equation~(\ref{eq:signal exact}) and (\ref{eq:noise exact}), it is
useful to define its the theoretical bounds independent of the
given form of the hypotheses. Comparing
equations~(\ref{eq:signal}) and (\ref{eq:noise}), we obtain
\begin{equation}\label{eq:signal to noise dist}
\frac{S_0^2}{2\sigma_N^2} \leq \frac{S_{\rm exp}^2}{N_{\rm exp}^2}
\leq \frac{S_0^2}{\sigma_N^2} + N^{\rm fid}+ N^{\rm test} ,
\end{equation}
where $S_0$ is given by equation~(\ref{eq:S_0}).  The high bound is
approached when $H^{\rm test}$ tends to be parallel to $H^{\rm fid}$,
the low bound is approached when $H^{\rm test}$ is orthogonal to
$H^{\rm fid}$ near $H^{\rm fid}(p^{\rm fid}_0)$ and $H^{\rm
test}(p^{\rm test}_0)$.  Here $H$ denotes the sub--space of the
K--dimensional vector space that is spanned by the variable parameters.


\section*{APPENDIX C: PARAMETER BIAS FOR A FALSE HYPOTHESIS}

Here we present an estimate for the values of the best--fitting
parameters of a false hypothesis. This problem can be contrasted
with the application of the Fisher matrix.  While the latter is a
method to obtain the variance on the parameters of the true model,
here we seek an approximation of the expectation values for the
best-fitting parameters of a false test model (regardless of
the uncertainties around this best fit, and whether the best
is an acceptable fit).

Finding $p^{\rm test}_0$, the minimum value in the $p^{\rm test}$
parameter space in general can be computationally tedious. Our
treatment is general as long as the fiducial hypotheses has all of
the parameters of the test hypotheses plus some additional
parameters and if the two hypotheses deviate by only a reasonably
small amount. In particular, as displayed in the panels of
Figure~\ref{fig:H2_minimum34}, the numerical value obtained naively
by
\begin{equation}
{p^{\rm test}_0}^{(0)}=(\Delta T^{\rm test}_s,c^{\rm
test})=(\Delta T^{\rm fid}_s,c^{\rm fid})
\end{equation}
is a fair zeroth approximation for the two cluster models in most
cases. A better approximation can be obtained by expanding
$[{\bf h}^{\rm test}(p^{\rm test})-{\bf h}^{\rm fid}(p^{\rm
fid})]^2$ to second order around ${p^{\rm test}_0}^{(0)}$ and
finding its minimum in terms of $p^{\rm test}$.
\begin{align}\nonumber
({\bf h}^{\rm test}(p^{\rm test}_0) &- {\bf h}^{\rm fid}(p^{\rm
fid}))^2\approx  \\
\label{eq:p' expansion} &({\bf h}^{\rm test}-{\bf h}^{\rm fid})^2 +
2(\partial_k{\bf h}^{\rm test})\cdot({\bf h}^{\rm test}-{\bf h}^{\rm
fid})
x_k  \\
&+ [(\partial_j{\bf h}^{\rm test})\cdot(\partial_k{\bf h}^{\rm
test}) +
\partial_j\partial_k {\bf h}^{\rm test}\cdot({\bf h}^{\rm test}-{\bf h}^{\rm fid})]
x_jx_k \nonumber
\end{align}
where it is assumed that ${\bf h}^{\rm test}$ and ${\bf h}^{\rm
fid}$ are evaluated at ${p^{\rm test}_0}^{(0)}$ and $p^{\rm fid}$,
respectively. The indices $j$ and $k$ run across the test model
parameters, and $x_k={p^{\rm test}_0}^{(1)}_k-{p^{\rm
test}_0}^{(0)}_k$, furthermore
\begin{equation}
\partial_k=\frac{\partial}
{\partial {p^{\rm test}}_k} .
\end{equation}
Let us denote the quadratic coefficient by
\begin{equation}
M_{kj}=(\partial_k {\bf h}^{\rm test})\cdot(\partial_j {\bf h}^{\rm
test}) + ({\bf h}^{\rm test}-{\bf h}^{\rm
fid})\cdot\partial_k\partial_j {\bf h}^{\rm test} .
\end{equation}
Equation (\ref{eq:p' expansion}) gives the deviation of the signal
to noise ratio if the test model parameters are modified from the
fiducial values. The expected signal to noise corresponds to the
minimum value in terms of $x_k$, at which the $x_k$ derivative
must vanish
\begin{align}\label{eq:p' expansion derivative}
\frac{\partial}{\partial x_k}[ h^{\rm test}(p^{\rm test}_0) &-
{\bf h}^{\rm fid}(p^{\rm fid})]^2\approx \\
&2(\partial_k{\bf h}^{\rm test})\cdot({\bf h}^{\rm test}-{\bf h}^{\rm
fid}) + 2M_{kj}x_j=0 .\nonumber
\end{align}
Since $x_k={p^{\rm test}_0}^{(1)}_k-{p^{\rm test}_0}^{(0)}_k$, the
solution of this equation gives the next approximation of $p^{\rm
test}_0$. Equation~(\ref{eq:p' expansion derivative}) is solved by
inverting the coefficient matrix. Thus
\begin{equation}\label{eq:p' approximation}
{p^{\rm test}_0}^{(1)}_k={p^{\rm test}_0}^{(0)}_k -
M_{kj}^{-1}(\partial_j {\bf h}^{\rm test})\cdot({\bf h}^{\rm
test}-{\bf h}^{\rm fid}) ,
\end{equation}
${p^{\rm test}_0}^{(1)}_k$ is the improved approximation of the
$k^\mathrm{th}$ parameter of ${p^{\rm test}_0}$. Note that the
$\Delta T_s$-dependence of the hypotheses is linear, implying
$\partial_{\Delta T_s}{\bf h}^{\rm test}(p^{\rm test})={\bf
h}^{\rm test}(c^{\rm test},\Delta T^{\rm test}_s=1)$ and
$\partial^2_{\Delta T_s}{\bf h}^{\rm test}(p^{\rm test})=0$ for
all $p^{\rm test}$. The $c^{\rm test}$ parameter derivatives are
to be calculated numerically using equation~(\ref{eq:T(x)}). In
conclusion, finding the minimum in equation~(\ref{eq:signal to
noise advanced}) is simplified to evaluating parameter derivatives
of the ${\bf h}^{\rm test}$ function at only ${p^{\rm
test}_0}^{(0)}$.

Figure \ref{fig:H2_minimum34} shows how the best fitting ${p^{\rm
test}_0}$ parameters are related to the naive choice $p^{\rm fid}$
when detecting virialization shocks with the SZ effect. The $S/N$
function has a minimum at ${p^{\rm test}_0}$. Finding the critical
point with equation~(\ref{eq:p' approximation}) yields the correct
global minimum for $(M_\vir,z,\alpha)=(10^{13}M_\odot,0.3,1.5)$
for any $(x_\mx,D)$ and $(10^{14}M_\odot,0.3,1.5)$ for any
$(x_\mx,D)$ unless $x_\mx\approx c$. However, for
$(M_\vir,z,\alpha,x_\mx)\approx(10^{14}M_\odot,0.3,1.5,c)$ there
are multiple local minima and saddle points. The ${p^{\rm
test}_0}={p^{\rm test}_0}^{(1)}$ approximation of
equation~(\ref{eq:p' approximation}) yields the local minima in
the vicinity, which is \textit{not} the global minimum. For
$(M_\vir,z,\alpha)\approx(10^{15}M_\odot,0.3,1.5)$ and arbitrary
$(x_\mx,D)$, the ${p^{\rm test}_0}^{(1)}$ value corresponds to the
saddle point. Again, the ${p^{\rm test}_0}={p^{\rm test}_0}^{(1)}$
approximation of equation~(\ref{eq:p' approximation}) breaks down.

The saddle points can be identified by calculating $\det M_{i,j}$
\begin{equation}
\det M_{kj}(p^{test}) \begin{array}{ll}
  > 0 & \text{iff $p^{test}$ is at a local minimum or maximum}\\
  < 0 & \text{iff $p^{test}$ is at a saddle point} \\
\end{array}
\end{equation}
The saddle point can be eluded if $\det M_{kj}$ is calculated and
a step is made towards ${p^{\rm test}_0}^{(1)}$ only if $\det
M_{kj}>0$. If it is negative, then a step is made ''downhill''
along the negative gradient $-2(\partial_k{\bf h}^{\rm
test})\cdot({\bf h}^{\rm test}-{\bf h}^{\rm fid})$.

Repeating equation~(\ref{eq:p' approximation}) or the downhill
steps gives a numerical efficient way of obtaining better and
better approximations of the best fit test parameter values near
the corresponding fiducial parameters. Further study shows that
this algorithm leads to the correct ${p^{\rm test}_0}$ value in
most cases. Only the fiducial parameters around
$p=(M_\vir,z,\alpha)\approx(10^{14}M_\odot,0.3,1.5)$, with
$(x_\mx,D)=(c,0.01c)$ or $(c,0.5c)$ have non-global minimum in the
close neighborhood of $p^{\rm fid}_0$.
In this case the best fitting test model parameters have to be
obtained by evaluating equation~(\ref{eq:S_0}) at multiple parameter
values, scanning the parameter space.

\end{document}